\documentclass[useAMS,usenatbib,usegraphicx]{mn2e}

\usepackage{amsmath, amssymb, graphics, graphicx, rotating, lscape}
\usepackage[hang,scriptsize]{subfigure}
\usepackage{hyperref}

\graphicspath{{./figs/}}

\newcommand{\mathsym}[1]{{}}

\newcommand{\tref}[1]{Table~\ref{#1}}
\newcommand{\eref}[1]{Eqn.~\ref{#1}}

\title[Complex RM Structure]
{Complex Faraday depth structure of Active Galactic Nuclei as revealed by broadband radio polarimetry}

\author[O'Sullivan et al.]{S.~P.~O'Sullivan$^{1}$, S.~Brown$^{1}$, T.~Robishaw$^{2}$, D.~H.~F.~M.~Schnitzeler$^{1}$, \and
N.~M.~McClure-Griffiths$^{1}$, I.~J.~Feain$^{1}$, A.~R.~Taylor$^{3}$, B.~M.~Gaensler$^{4}$, \and T.~L.~Landecker$^{2}$, L.~Harvey-Smith$^{1}$, E.~Carretti$^{1}$ \\ \\
$^{1}$CSIRO Astronomy and Space Science, ATNF, PO Box 76, Epping, NSW 1710, Australia.\\
$^{2}$National Research Council Canada, Herzberg Institute of Astrophysics, DRAO, Penticton, B.C., V2A 6K3, Canada\\
$^{3}$Department of Physics and Astronomy, University of Calgary, 2500 University Drive NW, Calgary, AB, T2N 1N4, Canada.\\
$^{4}$Sydney Institute for Astronomy, School of Physics, The University of Sydney, NSW 2006, Australia.\\
}
\begin{document}

\date{Accepted 2012 January 13. Received 2011 December 1; in original form 2011 October 2}
\pagerange{\pageref{firstpage}--\pageref{lastpage}} \pubyear{2012}
\maketitle
\label{firstpage}
\begin{abstract}
We present a detailed study of the Faraday depth structure of four bright ($>1$~Jy), 
strongly polarized, unresolved, radio-loud quasars. The Australia Telescope Compact 
Array (ATCA) was used to observe these sources with 2~GHz of instantaneous 
bandwidth from 1.1 to 3.1~GHz. 
This allowed us to spectrally resolve the polarization structure of spatially 
unresolved radio sources, and by fitting 
various Faraday rotation models to the data, 
we conclusively demonstrate that 
two of the sources cannot be described by a simple rotation measure (RM) 
component modified by depolarization from a foreground Faraday screen. 
Our results have important implications for using background extragalactic 
radio sources as probes of the 
Galactic and intergalactic 
magneto-ionic media as we show how RM 
estimations from narrow-bandwidth observations can give 
erroneous results in the presence of multiple interfering Faraday components. 
We postulate that the additional RM components arise from polarized structure in the 
compact inner regions of the radio source itself and not from polarized emission from 
Galactic or intergalactic foreground regions. We further suggest that this may contribute significantly to 
any RM time-variability seen in RM studies on these angular scales. 
Follow-up, high-sensitivity VLBI observations of these sources will directly test our predictions. 

\end{abstract}

\section{Introduction} 

Radio-loud Active Galactic Nuclei (AGN) eject powerful jets of relativistic plasma whose 
polarized, non-thermal synchrotron radiation can be used as a probe of the 
magneto-ionic material along the entire line of sight between us and the source of emission. 
Many studies have used these extragalactic background sources 
to study the strength and structure of magnetic fields in our Galaxy 
\citep[e.g.][]{brown2007, taylor2009, mao2010, vaneck2011, harveysmith2011}, other 
galaxies \citep[e.g.][]{gaensler2005, mao2008, feain2009} and in galaxy clusters 
\citep[e.g.][]{laing2008, bonafede2010, pizzo2011}. 
Future studies on new revolutionary instruments such as the Australian Square Kilometre 
Array Pathfinder (ASKAP) and the Square Kilometre Array (SKA) will rely on 
these background sources to probe the strength, structure and evolution of cosmic magnetism 
in unprecedented detail \citep[e.g.][]{beckgaensler2004, gaensler2009}. 

In this paper, we present a detailed study of the polarization and rotation measure (RM) 
properties of four bright, unresolved, strongly polarized, radio--loud AGN. 
Using the new Compact Array Broadband Backend (CABB) system \citep{cabbpaper} 
on the Australia Telescope Compact Array (ATCA), spectropolarimetric 
studies of these AGN were performed using 2~GHz of instantaneous bandwidth on the upgraded 
receiver system from 1.1 to 3.1~GHz\footnote{http://www.atnf.csiro.au/observers/memos/AT39.3\_128.pdf}.  
All-sky RM surveys such as the planned Polarization Sky 
Survey of the Universe's Magnetism (POSSUM) on ASKAP 
will measure the RMs of $\sim3$~million extragalactic radio 
sources over 30,000 deg$^2$ \citep{possum}. POSSUM will likely have 
300~MHz of instantaneous bandwidth covering the frequency range from 1130 to 1430 MHz. 
Proper interpretation of the results from this huge dataset will require 
extensive testing of the algorithms used to accurately extract the polarization and RM 
properties of individual sources. The ATCA is an ideal instrument for this process, 
whilst also providing new and unique insights into the Faraday depth structure of 
extragalactic sources due to its wide-bandwidth and high spectral resolution.

Following \cite{sokoloff1998}, and references therein, we define the complex linear polarization as 
\begin{equation} \label{complexP}
P = Q+iU = p I e^{2i \Psi}
\end{equation}

\noindent where $I$, $Q$, $U$ are the measured Stokes parameters and $\Psi$ is the observed 
polarization angle. 
We use the notation of \cite{farnsworth2011} in defining $q=Q/I$ and $u=U/I$, so that the 
measured magnitude of the degree of linear polarization is
\begin{equation} \label{p}
p=\sqrt{q^2+u^2}
\end{equation}
and the polarization angle is 
\begin{equation} \label{EVPA}
\Psi=\frac{1}{2}\arctan\frac{u}{q}
\end{equation}
Taking the fractional values decouples depolarization effects 
from simple spectral index effects in analysing the dependence of polarization with wavelength.  
It also minimises errors in the estimate of the RM using the RM synthesis 
technique \citep{bdb2005}. 

The observed polarization angle $\Psi$ is modified from its intrinsic 
value ($\Psi_0$) by the effect of Faraday rotation, caused by 
magneto-ionic material between the source of polarized emission and the telescope. 
If there are different regions of polarized emission sampled within a single resolution 
element 
then each of these regions will likely experience different amounts of Faraday rotation. 
Hence, to describe the Faraday rotation of a particular region of polarized emission 
we use the Faraday depth \citep{burn1966}
\begin{equation} \label{FaradayDepth}
\phi = 0.81 \int^{\rm telescope}_{\rm emission}{ n \bmath{B}\cdot d\bmath{l}} ~~{\rm rad~m}^{-2}
\end{equation}

\noindent where $n$ is the free electron density (in units of cm$^{-3}$), 
$\bmath{B}$ is the magnetic field (in $\mu$G) and $\bmath{l}$ is the distance 
along the line of sight (in parsecs). 
\cite{bdb2005} define a ``Faraday thin'' source as one in which 
$\lambda^2\Delta\phi\ll1$, and a ``Faraday thick'' source in cases where 
$\lambda^2\Delta\phi\gg1$ (where $\Delta\phi$ is the extent of the source in Faraday depth 
and $\lambda$ is the wavelength). 

In the simplest possible scenario, in which there is a background source of emission 
and only pure rotation due to a foreground magneto-ionic medium, then the 
Faraday depth is equal to the RM and we get 
\begin{equation} \label{RM}
\Psi=\Psi_0+{\rm RM}\,\lambda^2
\end{equation}

Depolarization from radio sources, where the degree of polarization decreases 
with increasing wavelength, is typically modelled as a single RM component with 
external Faraday dispersion \citep[e.g.][]{tribble1991, rossetti2008}. 
Our new wide-bandwidth data allow a detailed investigation of 
the case of multiple interfering RM components either along the line of sight or intrinsic to the 
source itself. Multiple RM components can cause both increases and/or decreases in $p(\lambda^2)$ 
with $\lambda^2$ as well as, but not always, deviations from a linear $\Psi(\lambda^2)$ behaviour. 
\cite{slysh1965} first employed a two component model to explain polarization measurements 
of Cygnus A while \cite{goldsteinreed1984} applied a similar model to 3C~27. 
A more recent study by \cite{law2011} showed that multiple RM components 
could be identified in extragalactic point sources using the RM synthesis technique 
on wide-band data from 1.0 to 2.0~GHz. 
\cite{farnsworth2011} highlighted the importance of describing both $\Psi(\lambda^2)$ and 
$p(\lambda^2)$ in determining the correct Faraday depth structure of extragalactic sources 
using a combination of data at 350~MHz and 1.4~GHz. 
Following on from this work, we conclusively show the effect of multiple RM components 
in two extragalactic sources by considering several different Faraday rotation models to 
simultaneously describe both $\Psi(\lambda^2)$ and $p(\lambda^2)$. 
 
In Section 2, we describe the observations, source selection and calibration 
process. Section 3 outlines the RM synthesis technique while Section 4 
describes the various polarization models employed as well as our method 
for discriminating between models. Section 5 presents our results for each 
source in order of increasing RM complexity. Section 6 discusses the 
implications of this work and we list our conclusions in Section 7. 
Throughout this paper, we assume a cosmology with H$_0 = 71$ km s$^{-1}$ Mpc$^{-1}$, 
$\Omega_M=0.27$ and~$\Omega_{\Lambda}=0.73$, and define the spectral index, 
$\alpha$, such that the observed flux density ($I$) at frequency $\nu$ follows the relation 
$I_{\nu}\propto\nu^{\rm{+}\alpha}$.

\section{Observations and Data Reduction}

\begin{table*}
 \caption{Observed sources (listed in order of increasing RM complexity)}
 \centering
   \begin{tabular}{ccccccccccccccc}
    \hline\hline
       (1) & (2) & (3) & (4) & (5) & (6) & (7) & (8) & (9) & (10) & (11) & (12) & (13) \\
      Source   & RA           & DEC        &     $l$         & $b$        & beam                     & pa   & $z$ &    Date       & $t_{\rm obs}$& $\lambda_0$ & $I_{\lambda_0}$  & $\alpha$ \\
Name         &  [J2000]  & [J2000]  & [$^\circ$]  & [$^\circ$]   &    [$" \times"$]       &   [$^\circ$] &          &                  &   [min]               &     [m]               &       [mJy]                &                   \\
      \hline            
PKS B1903-802  & 19:12:40.0 &  -80:10:05.9 & 314.0 & $-27.6$ & $11\times9$ & $-10$ & 0.50 & 2011-01-20 & 23                 & 0.141             & 1104                        & $-0.04$ \\
PKS B0454-810  & 04:50:05.4 &  -81:01:02.2 & 293.9 & $-31.4$ & $12\times9$ & $+60$ & 0.44 & 2011-01-20 & 16                 & 0.108             & 1009                        & $+0.46$ \\
PKS B1610-771  & 16:17:49.2 &  -77:17:18.5 & 313.4 & $-18.9$ & $10\times9$ & $+30$ & 1.71 & 2011-01-20 & 12                 & 0.137             & 3022                       & $-0.36$ \\
PKS B1039-47    & 10:41:44.6 &  -47:40:00.1 & 281.4 &   $+9.7$   & $12\times9$ & $0$ & 2.59 & 2011-01-09 & 22                        & 0.164             & 1699                        & $-0.38$ \\
      \hline
   \end{tabular}\\
\scriptsize{Column designation: 1 - Source name (IAU B1950.0); 2 - Right Ascension in J2000 coordinates; 3 - Declination in J2000 coordinates; 4 - Galactic longitude in degrees; 5 - Galactic latitude in degrees; 
                    6 - Synthesised beam in arcseconds; 7 - Synthesised beam position angle in degrees; 8 - redshift, taken from \cite{stickel1994}, and references therein, except for PKS B1039-47 (O. Titov, private communication); 9 - Date of observations; 10 - Total integration time in minutes; 
                    11 - Weighted mean $\lambda$; 12 - Total intensity at $\lambda_0$; 13 - Spectral index ($\alpha$), defined as $I_{\lambda}\propto\lambda^{-\alpha}$, calculated from our observations.}
\label{obs_params}
\end{table*}

\begin{figure}
    \includegraphics[width=6cm,angle=-90]{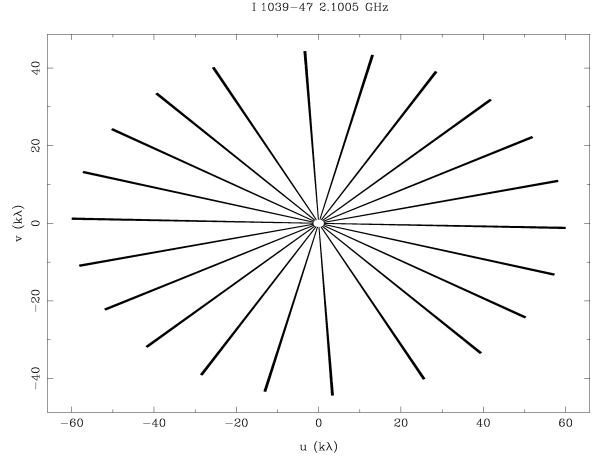}
      \caption{Plot showing the typical $uv$-coverage for the sources presented 
      in this paper (PKS B1039-47 shown). 
}
  \label{uvcoverage}
\end{figure}

The sources presented in this paper (\tref{obs_params}) were observed on 9 Jan 2011 and 20 Jan 2011 
with the ATCA from 1.1--3.1~GHz (with 1~MHz spectral resolution) in the 6A array configuration as part of a 
larger project to find suitable polarization calibrator sources for the ASKAP telescope. Figure~\ref{uvcoverage} 
shows the typical $uv$-coverage for sources in our experiment. 
The ATCA has six 22m antennas with linear feeds and a maximum baseline of 6~km providing 
an angular resolution of $\sim$10" at 1.4~GHz. It has recently undergone a major upgrade 
with the capabilities of the new CABB system described in detail by \cite{cabbpaper}. 

Since there are no 1.4~GHz polarization surveys of the southern sky 
at a resolution better than 36$^{\prime}$ \citep{testori2008}, we compiled a 
list of sources suitable for ASKAP polarization calibration from archival observations. 
Candidate sources were selected by searching the 
\href{http://atoa.atnf.csiro.au/}{ATCA online archive}\footnote{http://atoa.atnf.csiro.au/} for sources that 
had Stokes \emph{I} flux densities greater than 1 Jy at 843~MHz in the SUMSS 
catalogue \citep{mauch2003}. The extracted sources, which had been observed 
using the old narrow-band 128~MHz system, were then calibrated and imaged 
in \href{ftp://ftp.atnf.csiro.au/pub/software/miriad/userguide.ps.gz}{\textsc{Miriad}} 
using standard techniques \citep{sault1995}. 
Eight of the brightest sources found in polarized intensity were selected along with two 
unpolarized sources (fractional polarization $< 0.1$\%) 
for high precision polarization observations with the new wide-band CABB system on the ATCA. 
We selected four sources for detailed analysis in this paper that had reliable calibration 
across the full 2~GHz band and displayed a range of RM complexity from single to multiple 
RM components.  

In order to avoid any frequency-dependent calibration effects, the data for each source
were split up at 128~MHz intervals and calibrated using the \textsc{Miriad} 
software package. 
For the absolute flux density scale correction and bandpass calibration, we used a single 
observation of the ATCA primary flux calibrator PKS B1934-638 on each day. 
Flagging was done in an automated fashion (after bandpass calibration) using 
\href{http://www.atnf.csiro.au/people/Emil.Lenc/tools/Tools/Mirflag_Plotvis.html}{\textsc{Mirflag}} 
\citep{pieflag}. 
Some minor manual flagging was sometimes required afterwards. 
In total $\sim$$10-30\%$ of the 2~GHz band was lost due to radio-frequency interference (RFI), mainly between 1.1 
and 1.7~GHz. 
The leakage and complex gain solutions were determined for each individual 
128~MHz sub-band before recombining the entire band. 

One of our targets, PKS B1903-802, was used on 20 Jan 2011 as the secondary calibrator to correct for 
any atmospheric phase variations as well as the polarization leakages 
(21 x 1~min cuts, with one cut approximately once every hour, covering $340^{\circ}$ of parallactic angle). 
Because PKS B1903-802 is strongly polarized across the entire 2~GHz band, 
the $XY$ phase variations on the reference antenna could be determined and we were able 
to calibrate the absolute polarization position angle to within $\pm1^{\circ}$. 
These solutions were then copied to the other sources and amplitude and 
phase self-calibration was performed on PKS B0454-801 and PKS B1610-771. 
Our other target, PKS B1039-47, was used for polarization calibration on 9 Jan 2011 
(11 x 2~min cuts every hour, with a parallactic angle coverage of $210^{\circ}$).

Our calibration strategy allowed us to calculate the on-axis leakages in 128~MHz 
intervals. The results show that the real part of the leakage solution, which gives information 
about how the linear feeds deviate from perfect orthogonality (feed misalignment), is constant to 
better than 0.5\% across the entire band. The imaginary 
part, which probes the feed ellipticity (i.e.~how the feeds differ from perfectly linear feeds), 
has maximum deviations of up to $\sim2\%$ between 1.2 and 1.8~GHz. 
We find little variation between the leakage solutions on different days suggesting that the 
polarization performance of the wide-bandwidth system is stable on timescales of 
at least a few weeks.

\section{Extracting the Polarized Signal}
We first created uniformly-weighted $I$, $Q$ and $U$ images for each source in 10~MHz 
intervals and then deconvolved these maps using the H\"ogbom CLEAN algorithm \citep{hogbom1974}. 
In order to avoid any resolution dependent effects across the 2~GHz band, we smoothed 
each image to the resolution at the lowest frequency (see Table \ref{obs_params}). 
Since all sources were spatially unresolved, we took the emission at the position of the 
source in the Stokes $I$ image and created a table of $q$ and $u$ as a function of 
$\lambda^2$, at intervals in $\lambda^2$ corresponding to 10~MHz steps. Errors 
in each channel measurement were assigned using the rms noise from a small 
area around the source position in the clean-residual images. 

We then used the RM synthesis technique \citep{bdb2005} to extract the polarized signal 
over a wide range of possible Faraday depths. RM synthesis is a powerful analysis tool 
for polarimetry data since it helps overcome problems such 
as bandwidth depolarization and $n\pi$ polarization angle ambiguities. 
A spectrum of complex polarization versus Faraday depth was created from these 
data using the equation 
\begin{equation} \label{RMsynth}
F(\phi)= \sum^{N}_{j=1} w_j P_j e^{-2i\phi(\lambda_j^2-\lambda_0^2)}  \Big/   \sum^{N}_{j=1} w_j 
\end{equation}
\noindent where $N$ is the number of input maps, $P_j$ is the complex polarization 
at channel $j$ and $w_j$ are the weights (inverse square of the rms noise). Our reference 
wavelength ($\lambda_0$) is defined as 
\begin{equation} \label{lambdazero}
\lambda_0^2 =  \sum^{N}_{j=1} w_j \lambda^2_j  \Big/   \sum^{N}_{j=1} w_j .
\end{equation}
\noindent Essentially this derotates the $q$ and $u$ data for a particular assumed RM value and then 
sums the signal across the band; at the correct RM, the channels add coherently 
giving the maximum polarized intensity and the sensitivity of the full bandwidth. 

\begin{figure}
    \includegraphics[width=8cm]{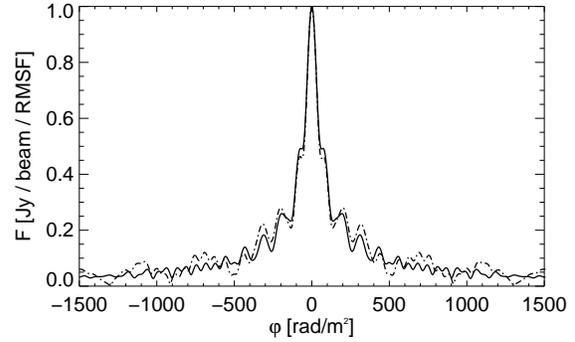}
      \caption{Plots of the rotation measure spread function (RMSF). 
      Solid line: RMSF from Jan 20. Dash-dot line: RMSF from Jan 9. 
}
  \label{RMSF}
\end{figure}

Figure~\ref{RMSF} shows the Rotation Measure Spread Function (RMSF) for the observations 
on both days. The RMSF is the normalised response function in Faraday depth space 
to the incomplete $\lambda^2$-sampling (i.e. with perfect $\lambda^2$ coverage this 
would be a delta function). More data were flagged on 9 Jan 2011 than on 20 Jan 2011, so that 
the RMSF for 9 Jan has slightly stronger sidelobes than on 20 Jan. 
\tref{CABB_RM} lists, for our observations, the Faraday depth resolution 
($\delta\phi\approx2\sqrt{3} / \Delta(\lambda^2)$, i.e. the FWHM of the RMSF), 
the largest detectable scale in Faraday depth space 
(max-scale $\approx\pi\lambda_{\rm min}^{-2}$) and 
the maximum observable Faraday depth ($\phi_{\rm max}\approx\sqrt{3}/\delta\lambda^2$), 
where $\Delta(\lambda^2)$ is the total bandwidth in $\lambda^2$-space 
(i.e. $\Delta(\lambda^2)=\lambda^2_{\rm max}-\lambda^2_{\rm min}$) 
and $\delta\lambda^2$ is the channel width. 

For each source, we initially searched for polarized power from 
$\pm\phi_{max}$ at Faraday depth intervals of 10~rad~m$^{-2}$ 
which corresponds to $\sim$6 Faraday depth intervals per $\delta\phi$ (see Table \ref{CABB_RM}). 
No significant power was found at large values of $|\phi|$. 
For the rest of our analysis we restricted our range to $\pm1000$~rad~m$^{-2}$ at 
1~rad~m$^{-2}$ intervals ($\sim60$ Faraday depth intervals per $\delta\phi$). 
We use RMCLEAN, as described by \cite{heald2009}, to deconvolve the ``dirty'' RM 
spectrum in an attempt to recover information lost due to the incomplete frequency coverage. 
For each source, we cleaned down to the rms noise level ($\sigma_{q,u}$) listed in Table~3.
In principle, the RM synthesis technique should be able to detect regions 
that are extended in Faraday depth space as long as the region extends 
beyond the FWHM of the RMSF. 

\begin{table}
  \caption{RM Synthesis Capabilities of 1.1--3.1~GHz ATCA band}
    \centering
    \begin{tabular}{cccccccccccccc}
    \hline\hline
    (1) & (2) & (3) & (4) & (5) \\
    Frequency & Resolution  & $\delta \phi$ & $\textnormal{max-scale}$ & $|\phi_{max}|$ \\    
 ${\rm [GHz]}$ & [MHz] & [rad m$^{-2}$] & [rad m$^{-2}$] & [rad m$^{-2}$]\\
    \hline
    1.1--3.1   & 1.0  & 60 & 340  & 13000  \\
    \hline
    \end{tabular}\\
    \scriptsize{Column designations: 1 - Instantaneous frequency coverage in GHz; 
    2 - Spectral resolution in MHz; 3 - Resolution in Faraday depth space; 
    4 - maximum detectable Faraday thickness; 5 - Maximum detectable Faraday depth.}
  \label{CABB_RM}
\end{table}

\section{Modelling Procedure}
To model the polarized signal in the presence of 
Faraday rotation in the simplest case, we use the equation 
\begin{equation} \label{simple}
P=p_0 e^{2i(\Psi_0+{\rm RM}\lambda^2)}
\end{equation}
where $p_0$ is the intrinsic degree of polarization of the synchrotron 
emission, 
$\Psi_0$ is the intrinsic polarization angle at the source of the emission and 
the RM describes the Faraday rotation caused by the foreground 
magneto-ionic material with the sign indicating whether the line-of-sight 
magnetic field is pointing towards us (positive RM) or away from us (negative RM). 
The data for all our sources show changes in the degree of polarization 
across the observed wavelength range so we now consider 
the possible mechanisms behind this effect. 

Depolarization towards longer wavelengths can occur 
due to mixing of the emitting and rotating media, as well as 
from the finite spatial resolution of our observations. 
There are three commonly listed depolarization mechanisms 
(see \cite{sokoloff1998} for more detailed discussion of 
each case). 

1.~Differential Faraday rotation (DFR): this occurs when the emitting 
and rotating regions are co-spatial and are in the presence of a 
regular magnetic field. 
The polarization plane of the emission at the far side of the region 
undergoes a different amount of Faraday rotation compared to the 
polarized emission coming from the near side, causing depolarization when summed over the entire region. 
For the particular case of a uniform slab we have 
\begin{equation} \label{DFR}
P=p_0 \frac{\sin {R} \lambda^2}{ {R} \lambda^2} e^{2i(\Psi_0+\frac{1}{2}{R}\lambda^2)}
\end{equation}
\noindent where $R$ is the Faraday depth through the region.

2.~Internal Faraday dispersion (IFD): this occurs when the emitting and rotating regions also contain 
a turbulent magnetic field. In this case depolarization occurs because the plane of polarization 
experiences a random walk through the region. 
For identical distributions of all the constituents of the magneto-ionic medium along the line of sight, 
it can be described by
\begin{equation} \label{IFD}
P=p_0 e^{2i\Psi_0} \left( \frac{ 1-e^{2i{R}\lambda^2-2\varsigma^2_{\rm RM}\lambda^4} }{ 2\varsigma^2_{\rm RM}\lambda^4-2i{R}\lambda^2 } \right)
\end{equation}
where, in this case, $\Psi_0=\pi/2$ for a purely random anisotropic magnetic field and 
$\varsigma_{\rm RM}$ is the internal Faraday dispersion of the random field. 

3.~External Faraday dispersion/beam depolarization: this occurs in a purely external, non-emitting Faraday 
screen. In the case of turbulent magnetic fields, depolarization occurs when many turbulent 
cells are within the synthesised telescope beam\footnote{However, see \cite{tribble1991} for a detailed description of what 
happens when this assumption does not hold.}. On the other hand, for a regular magnetic field, any variation in the 
strength or direction of the field within the observing beam will lead to depolarization. 
Both effects can be described by
\begin{equation} \label{EFD}
P=p_0 e^{-2\sigma^2_{\rm RM}\lambda^4} e^{2i(\Psi_0+{\rm RM}\lambda^2)}
\end{equation}
where $\sigma_{\rm RM}$ is the dispersion about the mean RM across the source on the sky. 

A fourth possibility is a changing degree of polarization due to multiple interfering RM components, 
either along the line of sight or on the plane of the sky on scales smaller than our spatial resolution. 

To model our data, we simultaneously fit both the $q(\lambda^2)$ and $u(\lambda^2)$ data to the 
different polarization models listed above. 
We first tried a simple one-component RM model, which cannot 
describe any variation in the degree of polarization. We then tried to account for changes in $p(\lambda^2)$
by fitting a single RM component plus depolarization model, 
and in cases where this could not adequately describe the data, we then tried multiple RM-component models. 
In order to limit the number of models to investigate, we mainly considered models of 
either solely Faraday thin components (e.g.~emission from the radio galaxy only) or models of one 
Faraday thick component (e.g.~Galactic slab or mixed emitting and rotating region in the source) plus 
Faraday thin component(s). We believe that these models are the most physically reasonable 
cases for spatially unresolved extragalactic radio sources. 
However, we do not consider 
differing spectral indices of individual components. 
Hence, multiple component models are simply constructed as 
$P=P_1+P_2+\dots+P_N$.

\subsection{Model-fit Evaluation}
We utilised the maximum likelihood method to find the best-fit model parameters. 
The results from RM synthesis were used as a guide to our initial guesses for the 
RM and fractional polarization values. 
Each data 
point in the fit was weighted by the inverse square of the rms noise from the clean-residual image. 

The likelihood is the probability of obtaining the data, $d$, given a model of 
the source and some characterisation of the noise. Our data in this context 
are $q(\lambda^{2})$ and $u(\lambda^{2})$. For example, in the one-component 
model, we adopt $q_{model,i}=p_0\cos(2\Psi_{0}+2{\rm RM}\lambda_i^{2})$ and 
$u_{model,i}=p_0\sin(2\Psi_{0}+2{\rm RM}\lambda_i^{2})$, and we assume that 
$q_{i}=q_{model,i}+n_{i}$, where $n_{i}$ is Gaussian noise for channel $i$. 
The prior likelihood of a particular RM value for an observation of a 
single channel $i$ under the assumption of Gaussian noise is 

\begin{equation} \label{prior}
P_{i}(d_{i}|{\rm RM})=\frac{1}{\pi \sigma_{q_{i}} \sigma_{u_{i}}} \exp{\left(-\frac{(q_{i}-q_{model,i})^2}{2\sigma_{q_{i}}^{2}}-\frac{(u_{i}-u_{model,i})^2}{2\sigma_{u_{i}}^{2}}\right)}
\end{equation}

\noindent where $\sigma_{q, u}$ is the single channel rms. If we have 
a total of $N$ channels, the prior likelihood is now 

\begin{equation} \label{likelihood}
P(d|{\rm RM})=\prod_{i=1}^{N} P_{i}(d_{i}|{\rm RM}) 
\end{equation}

We used the \emph{Mathematica}\footnote{Wolfram Research, Inc., Mathematica, Version 7.0, Champaign, IL (2008).} 
function \textsc{Nonlinearmodelfit} to find the 
maximum of \eref{likelihood}, $L\equiv max(P(d|{\rm RM}))$. 
In the case of multiple models giving good fits to the data, we then used the 
Bayesian Information Criterion (BIC) to distinguish the goodness-of-fit between 
different models with different degrees-of-freedom \citep{BIC, BICastro};  

\begin{equation} \label{BIC}
{\rm BIC} \equiv -2\log P(d|k)\approx -2 \log L + k \log N
\end{equation}
\noindent where $k$ is the number of free parameters in the model. 
Hence, models with more parameters are heavily penalised given the 
large number of data points. 
We consider ${\rm BIC_{model_1}}-{\rm BIC_{model_2}} > 100$ to 
significantly favour model 2 (at 99\% level) across the measured parameter space. 
Bayesian model comparisons require an alternative model against 
which the comparison is made (i.e. a model cannot be rejected unless 
an alternative explanation is available that better fits the observations). 
In order to give a quantitative measure of how each individual 
model fits the data, we also calculate the reduced chi-square ($\chi^2_{\nu}$) 
goodness-of-fit values, where $\chi^2_{\nu}$ is obtained by dividing the sum of 
squared residuals by $\sigma_{q,u}^2$ and the number of degrees of freedom. 

\section{Results}
We present the sources in order of RM complexity (as listed in 
Table~\ref{obs_params}) with PKS B1903-802 having the simplest 
RM structure and PKS B1039-47 having the most complex. 
All models with their best-fit parameters for each source are 
listed in Table~3 (with the most-likely model highlighted in bold). 
The associated errors of each parameter are formal fitting errors 
which are calculated from the square root of the 
estimated error variance of each parameter. These 
errors have little meaning when the incorrect model is applied. 
Note that all polarized components listed are found with 
high significance. For example, the weakest 
polarized model component of 0.6\% listed in Table~3 has a 
signal to noise ratio (SNR) of $\sim$100. 

\begin{figure*}
  \subfigure[PKS B1903-802]{
    \includegraphics[width=7.5cm]{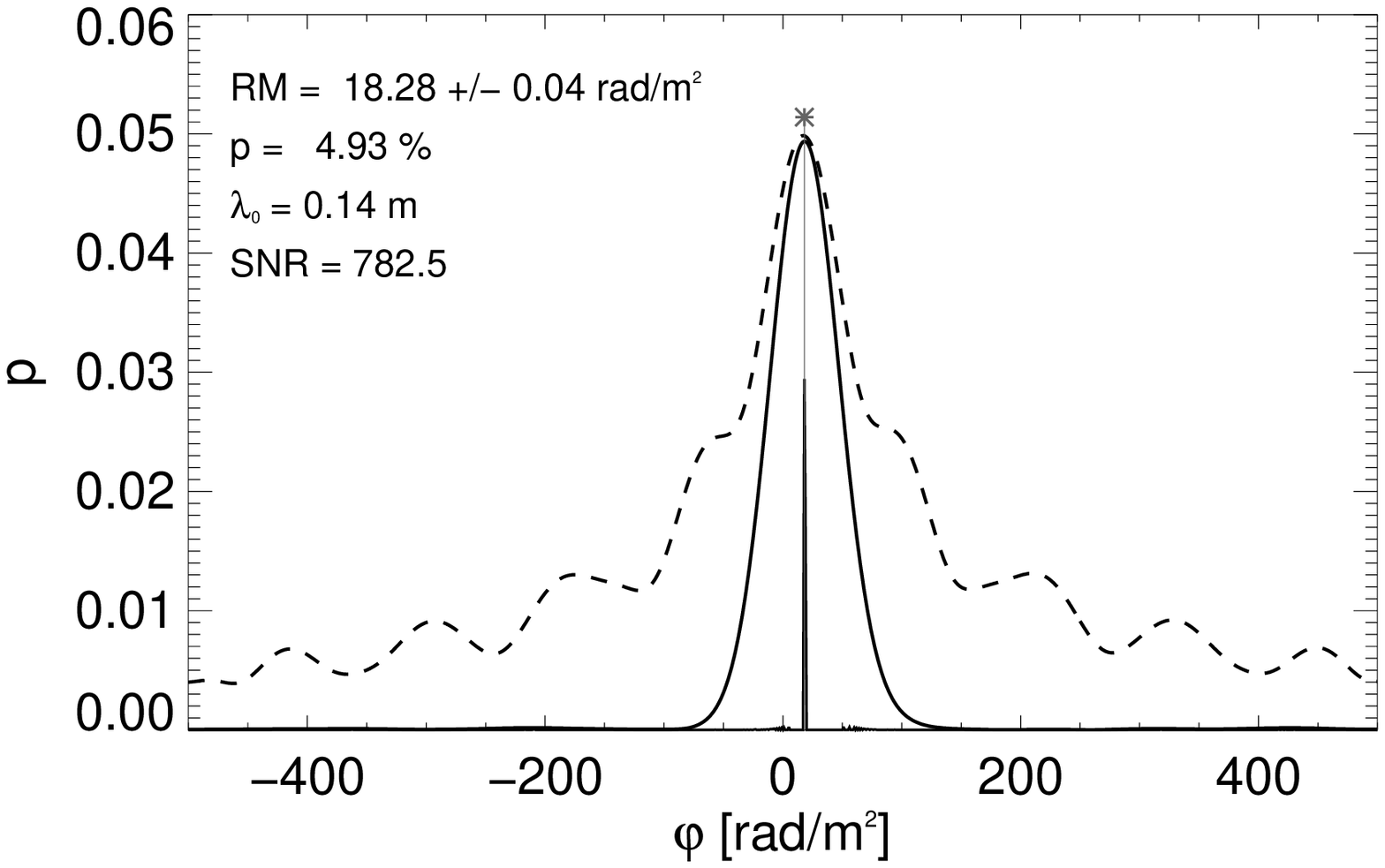}
    \label{Figure:figsubex:left}
  }
  \subfigure[PKS B0454-810]{
    \includegraphics[width=7.5cm]{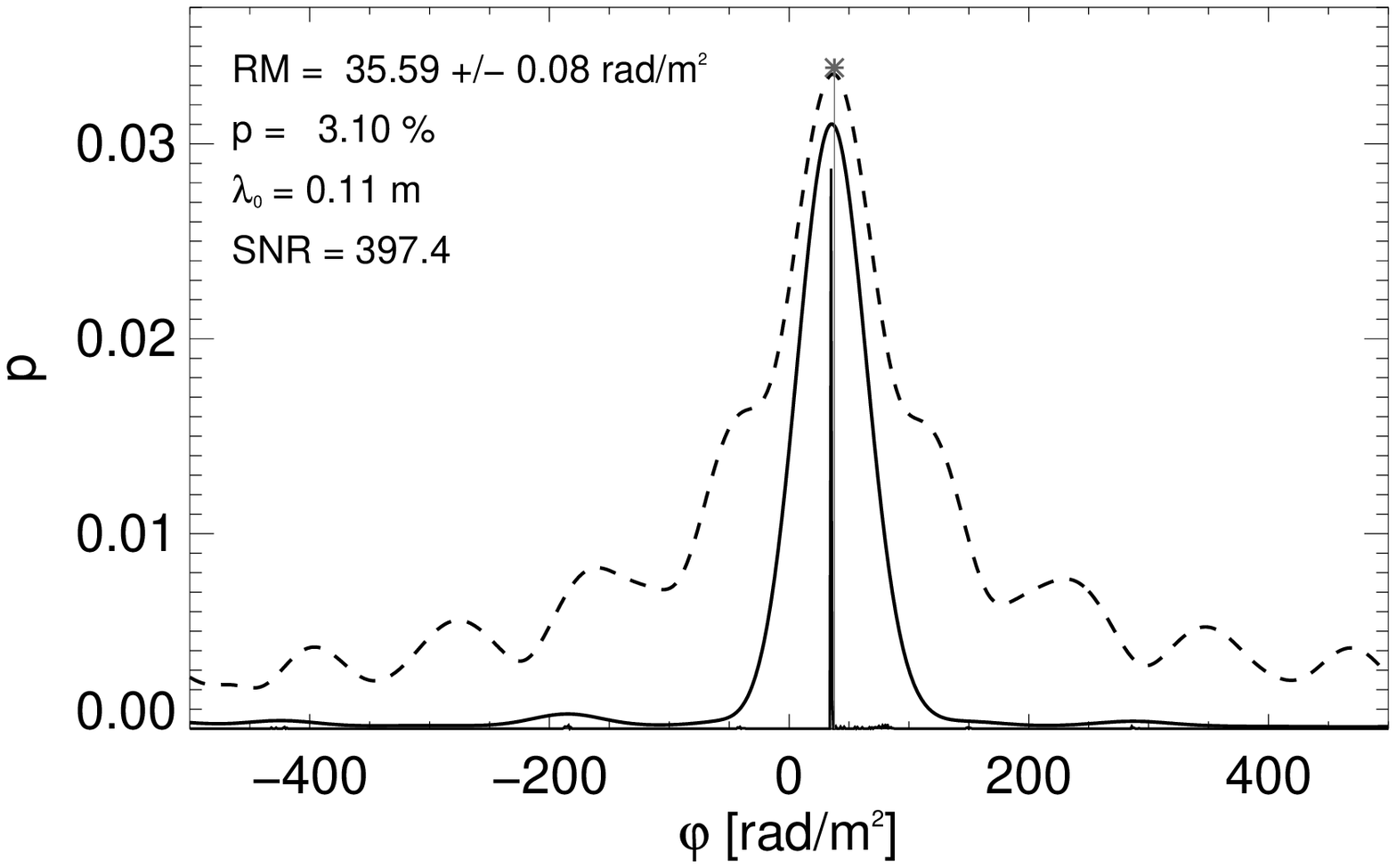}
    \label{Figure:figsubex:right}
  }
  \subfigure[PKS B1610-771]{
    \includegraphics[width=7.5cm]{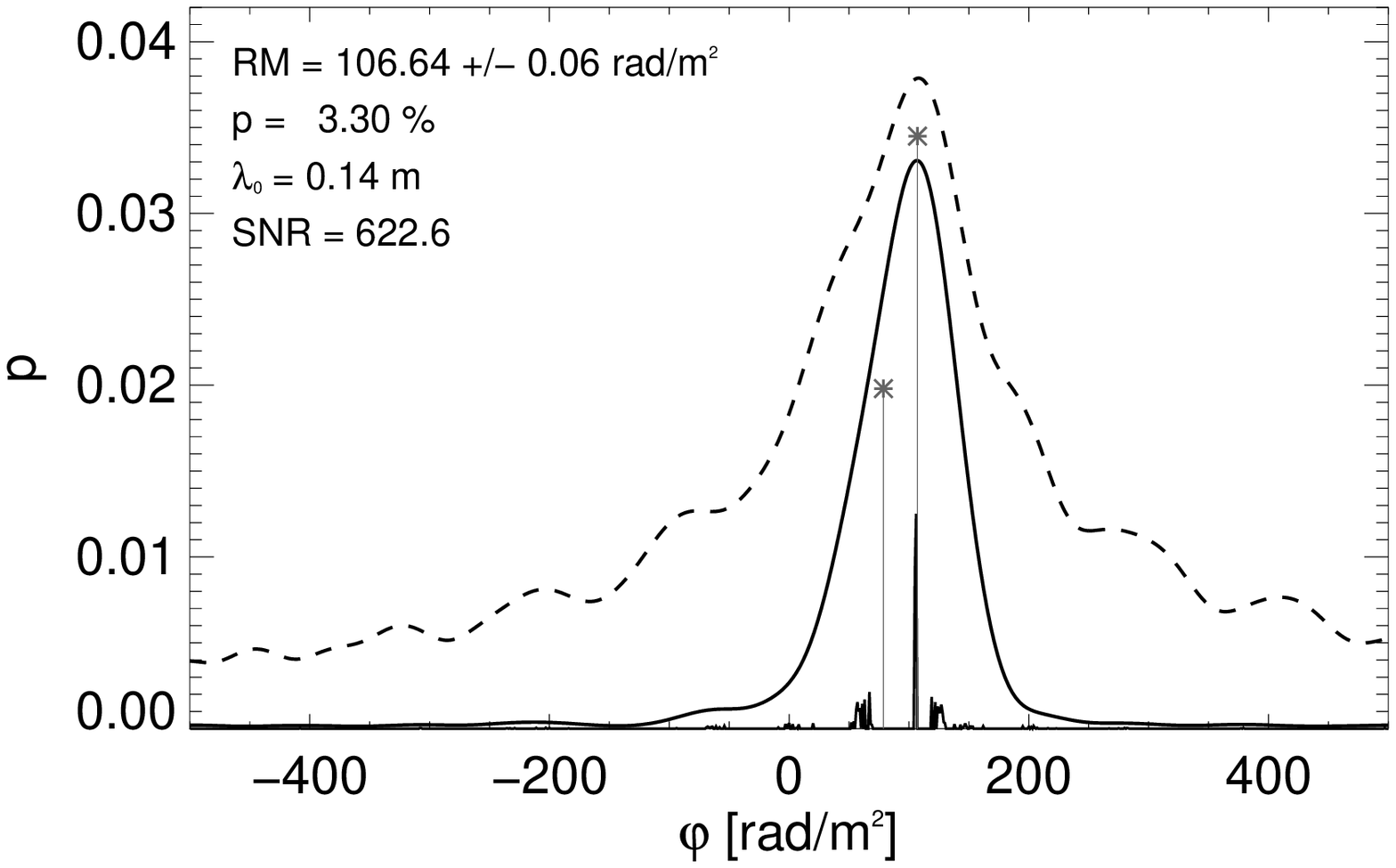}
    \label{Figure:figsubex:left}
  }
  \subfigure[PKS B1039-47]{
    \includegraphics[width=7.5cm]{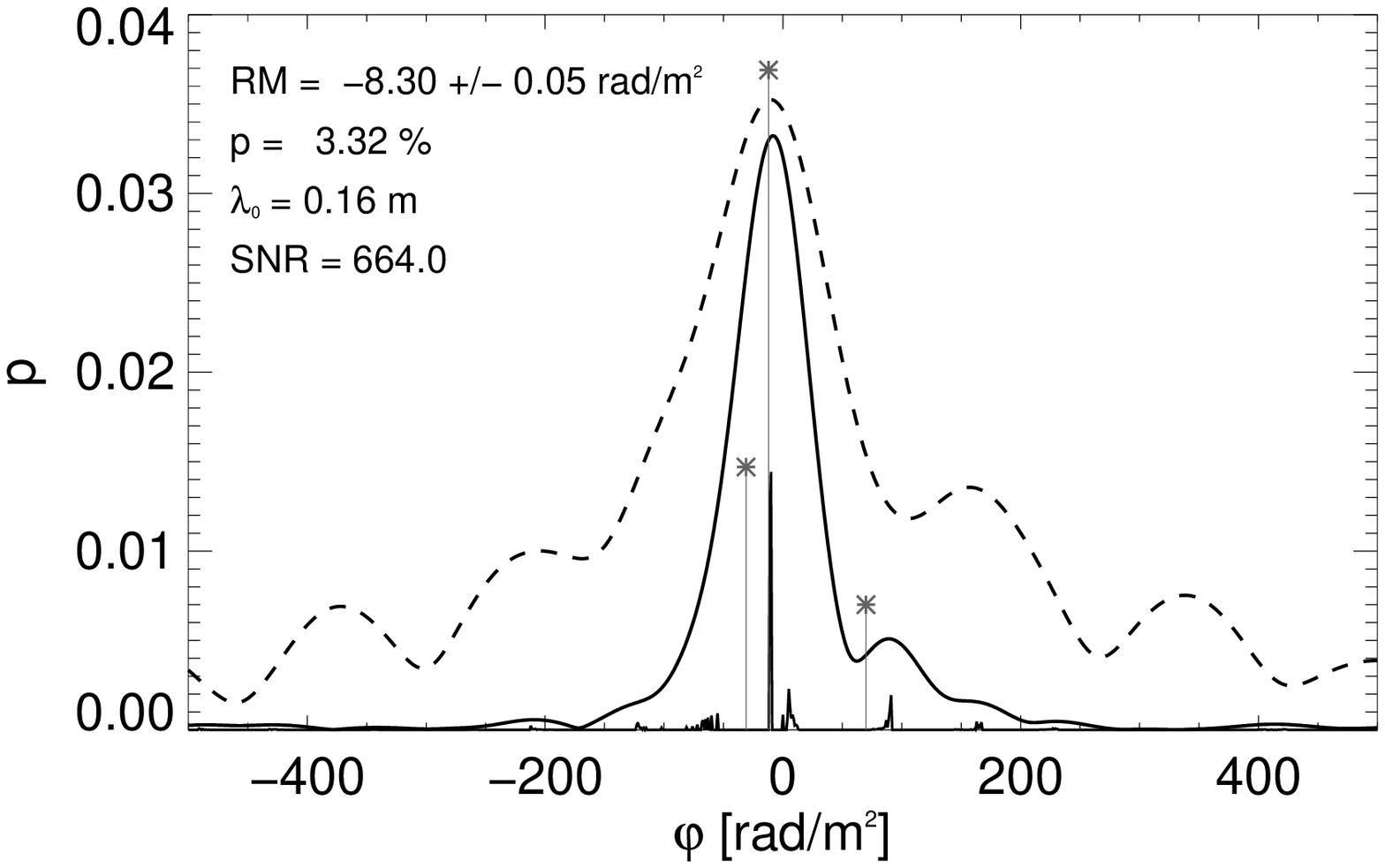}
    \label{Figure:figsubex:right}
  }  
  \caption{Plots of the RM spectra for all four sources. Dashed line: RM synthesis spectrum. 
  Solid line: RMCLEAN spectrum with the locations of the clean-components also shown as 
  vertical lines. The text in the top-left corner of each panel lists the parameters extracted 
  from the peak in the RMCLEAN spectrum as well as the value of the reference wavelength, $\lambda_0$. 
  The asterisks connected to the Faraday depth axis by the solid line denote the locations of 
  the best-fit model RM. 
  }
  \label{RMCLEAN}
\end{figure*}

In Figure~\ref{RMCLEAN}, we show the results of RM synthesis and RMCLEAN 
for all four sources. For each source we list the RM at the peak 
degree of polarization with its associated error. 
The uncertainty in the peak RM is calculated as the FWHM of the 
RMSF divided by twice the SNR \citep{bdb2005}. 
So for example, if we have an SNR of 600 and an RM resolution 
of $66$~rad~m$^{-2}$, then the quoted uncertainty in the peak RM 
is $0.06$~rad~m$^{-2}$. However, as described by \cite{law2011}, 
the accuracy of any individual RM-component value cannot be specified 
to better than the RM resolution due to the uncertainty in the 
distribution of components within the RM beam. 
The mean-weighted reference wavelength ($\lambda_0$) 
changes from source to source mainly due to different amounts 
of flagged data across the band for each source.

Faraday depth spectra for both PKS B1903-802 (Fig.~\ref{RMCLEAN}a) and 
PKS B0454-810 (Fig.~\ref{RMCLEAN}b) appear to be broadly consistent with a single RM component. 
The asymmetric distribution about the peak for PKS B1610-771, 
as well as the distribution of clean-components, indicates the presence 
of more than one RM component (Fig.~\ref{RMCLEAN}c). PKS B1039-47 has a 
distinct secondary peak at $\sim$100~rad~m$^{-2}$ and the 
clean-component locations suggest the presence of three or more 
RM components (Fig.~\ref{RMCLEAN}d) . We now discuss the Faraday rotation 
model-fits to the $q(\lambda^2)$, $u(\lambda^2)$ data for each source 
which are completely independent of the RMCLEAN results.

\subsection{PKS B1903-802}
\begin{figure*}
    \includegraphics[width=14cm]{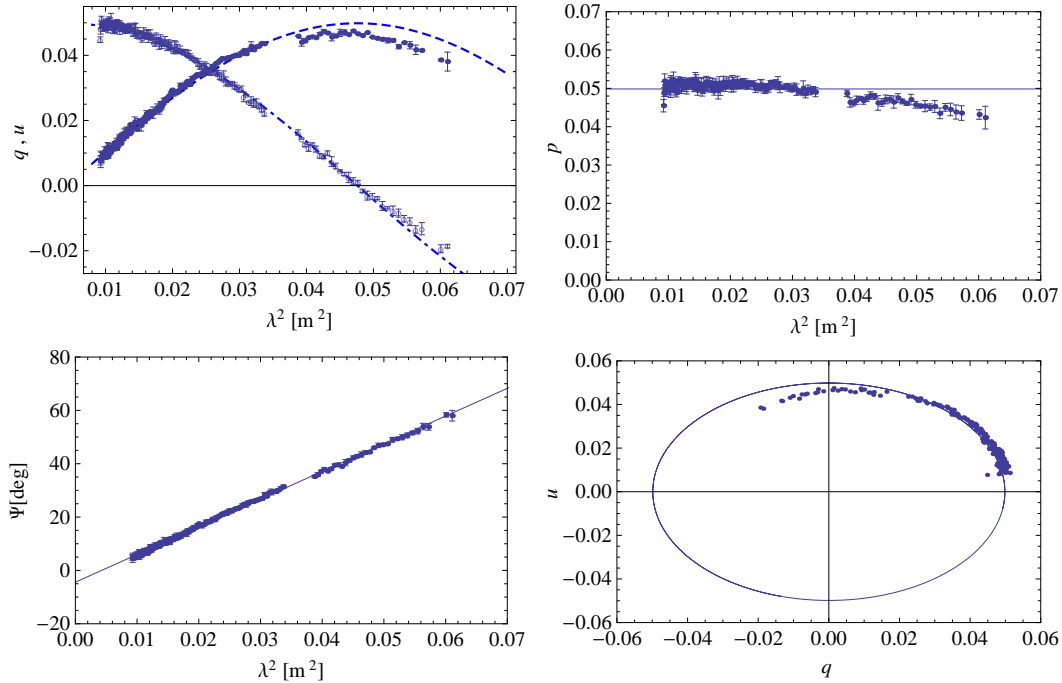}
      \caption{Polarization data for PKS B1903-802, and the corresponding 
      best-fit single RM-component model. (Eqn.~\ref{simple}). 
  Top left: $q$ (open circles) and $u$ (full circles) data vs.~$\lambda^2$, 
  fitted with the model $q$ (dot-dashed line) and $u$ (dashed line). 
  Top right: $p$ vs.~$\lambda^2$ data over-plotted by the model (solid line). 
  Bottom left: $\Psi$ vs.~$\lambda^2$ data over-plotted by the model (solid line). 
  Bottom right: $u$ vs.~$q$ data over-plotted by the model (solid line). 
}
  \label{1903simple}
\end{figure*}

\begin{figure*}
    \includegraphics[width=14cm]{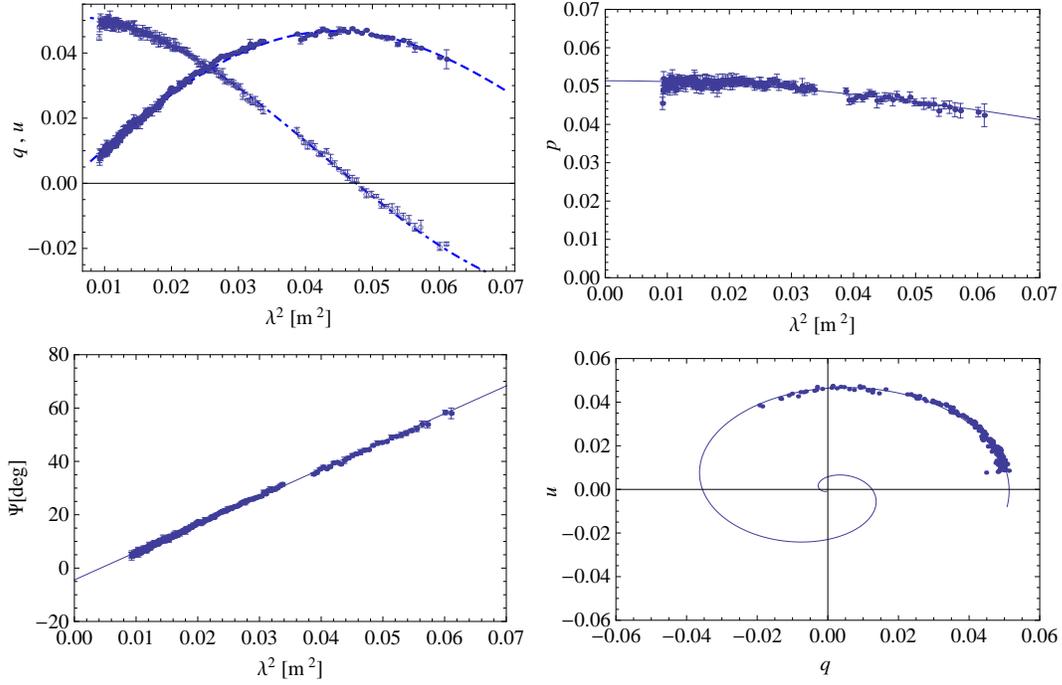}
  \caption{As for Figure~\ref{1903simple} for PKS B1903-802, but 
  now modelled by a single RM-component model with depolarization 
  from external Faraday dispersion (Eqn.~\ref{EFD}). 
}
  \label{1903depol}
\end{figure*}

\begin{figure*}
    \includegraphics[width=14cm]{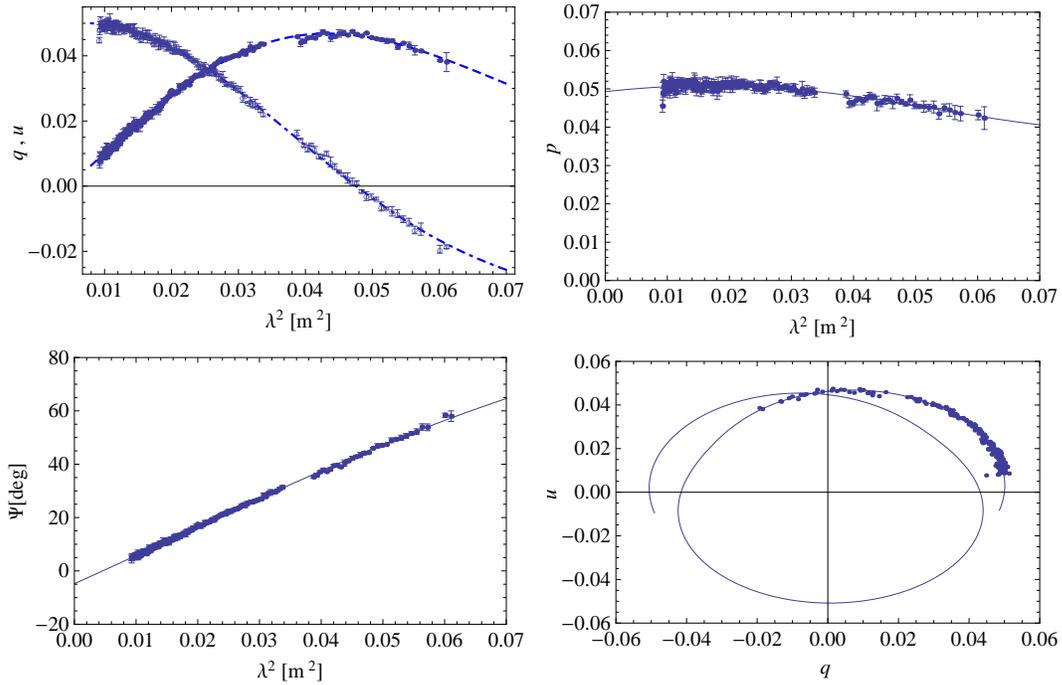}
  \caption{As for Figure~\ref{1903simple} for PKS B1903-802, 
  but fit with a two RM-component model. 
}
  \label{1903_2cmpnt}
\end{figure*}

A simple RM fit (Eqn.~\ref{simple}), as shown in Figure~\ref{1903simple}, provides a reasonable 
description of the $\Psi(\lambda^2)$ data but the $p$ vs.~$\lambda^2$ 
data clearly deviates from a constant degree of polarization. 
An external Faraday dispersion model (Eqn.~\ref{EFD}) provides an excellent fit 
to data, with our best fit model, shown in Figure~\ref{1903depol}, giving a polarized intensity 
of $5.14\pm0.04\%$, with a foreground RM of $+18.1\pm0.1$~rad~m$^{-2}$ and an 
external dispersion in RM across the source of $4.7\pm0.1$~rad~m$^{-2}$. 
However, in Figure~\ref{1903_2cmpnt} we can see that a two RM-component model also provides a very 
good description of the data with an RM of $+16.1\pm0.3$~rad~m$^{-2}$ for the stronger 
polarized component ($\sim$$4.5\%$) and an RM of 39.3$\pm2.0$~rad~m$^{-2}$ 
for the second polarized component ($\sim$$0.6\%$). 
This supports the conclusion of \cite{farnsworth2011} that modelling of both polarization amplitude 
and polarization angle is required in studies of Faraday rotation. 

In this case, neither the BIC nor the $\chi^2_{\nu}$ help us to clearly discriminate between models.  
The two models do not differ significantly over the measured parameter space so we cannot state with 
confidence which one is correct, although we favour the simpler external Faraday dispersion model. 
Lower frequency observations, from 700 MHz to 1 GHz on ASKAP for example, would 
clearly discriminate between the two models since for the two RM-component model 
there is a departure from a linear $\Psi$ vs.~$\lambda^2$ relationship over this 
range. The difference can be seen most clearly in the $q(\lambda^2)$ vs.~$u(\lambda^2)$ plots 
(i.e. compare Fig.~\ref{1903depol} and Fig.~\ref{1903_2cmpnt}). 

\begin{sidewaystable*}\label{modeltable} 
\scriptsize{
\centering
\vspace{-15cm} 
   \begin{tabular}{ccccccccccccccc}
   {\bf \small{Table 3.}} \\
    \hline\hline
       (1) & (2) & (3) & (4) & (5) & (6) & (7) & (8) & (9) & (10) & (11) & (12) & (13) & (14) & (15) \\    
     Source & Model & RM$_1$    & $p_{01}$ & $\Psi_{01}$ & RM$_2$    & $p_{02}$ & $\Psi_{02}$ & RM$_3$ & $p_{03}$ & $\Psi_{03}$ & $\sigma_{\rm RM}$ & $\sigma_{q,u}$ & $\chi^2_{\nu}$ & BIC \\
                   &             & [rad~m$^{-2}$] & [\%]          & [$^\circ$]       & [rad~m$^{-2}$] & [\%]            & [$^\circ$]     & [rad~m$^{-2}$] & [\%]           & [$^\circ$]      & [rad~m$^{-2}$]                 &    [\%]    &                            &         \\
    \hline
     PKS B1903-802 & Single RM, no screen                 & $+18.1(1)$ & $4.98(1)$   &  $-4.4(1)$ & -                        & -                          & -                      &   -           & -                 & -            &  -                         & 0.006            & 1.23 & $-3470$\\
                          & \bf{Single RM, foreground screen}  & $+18.1(1)$ & $5.14(4)$   &  $-4.4(1)$ & -                        & -                          & -                      & -            & -                & -            &   $4.7(1)$       & -                     & 1.03 & $-3714$\\
                          & 2 RM Components, no screen & $+16.1(3)$ & $4.50(8)$   &  $-2.4(3)$ & $+39(2)$ & $0.59(8)$ & $-25(2)$ & -        & -                & -            &      -                      & -                     &  1.04 & $-3779$\\
      \hline
     PKS B0454-810 & Single RM, no screen                 & $+37.9(2)$  & $3.36(2)$   &  $-48.6(3)$ & -                        & -                          & -                      &   -           & -                & -            &  -                           & 0.008           & 2.13 & $-3250$\\
                          & Single RM, foreground screen  & $+37.8(2)$   & $3.39(2)$  &  $-48.6(3)$ & -                        & -                          & -                      & -             & -                 & -            &    $2.8(6)$    & -                     & 1.60 & $-3250$\\
                          & 2 RM Components, no screen & $+29(1)$   & $2.7(2)$   &  $-33(2)$ & $+62(3)$   & $1.1(2)$ & $+88(5)$    & -         & -                 & -            &     -                       & -                      &  1.07 & $-3548$\\
      \hline
     PKS B1610-771 & Single RM, no screen                 & $+109.0(7)$ & $3.79(7)$   &  $+69(1)$    & -                        & -                          & -                      &     -         & -                & -            &  -                           &  0.005         & 97.3 & $-2346$\\
                          & Single RM, foreground screen  & $+104.3(4)$ & $5.20(4)$   &  $+73.9(5)$ & -                        & -                          & -                      & -             & -                 & -            &  $17.3$~(0.3)     & -                    & 1.41 & $-3024$\\
                          & \bf{2 RM Components, no screen} & $+107.1(2)$ & $3.45(4)$   &  $+83.3(6)$ & $+78.7(4)$   & $1.98(4)$    & $+73(1)$ & -         & -                 & -            &      -                       & -                     &  1.04 & $-3845$\\
      \hline
     PKS B1039-47   & Single RM, no screen                 & $-12.3(5)$ & $3.43(5)$   &  $+18.8(8)$ & -                        & -                          & -                      &    -          & -                & -            &  -                           & 0.005             & 127 & $-2362$\\
                          & 2 RM Components, no screen  & $-9.8(4)$  & $3.64(4)$    &  $+14.3(7)$ & $+85(2)$  & $0.68(4)$   & $0(4)$ & -             & -                 & -            &    -                   & -                     & 14.3 & $-2524$\\
                          & \bf{3 RM Components, no screen} & $-13(1)$ & $3.9(4)$   &  $+31(4)$  & $-30(2)$   & $1.7(4)$ & $0(8)$   & $+68(2)$  & $0.7(2)$ & $+35(3)$   & -     & -    &  1.23 & $-3194$\\
                          & 3 RM Components (1 DFR) & $-13(4)$ & $2.3(3)$  &  $+37(5)$ & $+70(1)$   & $0.7(2)$ & $+33(3)$  & $-21(4)$  & $2.9(1)$ & $+11(15)$  & -  & -     &  1.25 & $-3192$\\
      \hline 
   \end{tabular}
       \scriptsize{\\Column designation: 1 - Source name; 2 - Description of Faraday rotation model used; 3 - RM of first component ; 
       			4 - Degree of polarization of first component; 5 - Intrinsic polarization angle (at $\lambda=0$) of first component; 
                            6 - RM of second component ; 7 - Degree of polarization of second component; 8 - Intrinsic polarization angle of second component; 
                            9 - RM of third component ; 10 - Degree of polarization of third component; 11 - Intrinsic polarization angle of third component; 
                            12 - Dispersion about mean RM; 13 - rms noise level in fractional polarization; 14 -  Reduced chi-square goodness-of-fit value; 15 - Bayesian information criterion for model comparison. 
                            The favoured model for each source is indicated by bold-face font. The error in the final digit of all parameters is indicated by the number in the parentheses. 
                            }
}
\end{sidewaystable*} 


\subsection{PKS B0454-810}
\begin{figure*}
    \includegraphics[width=14cm]{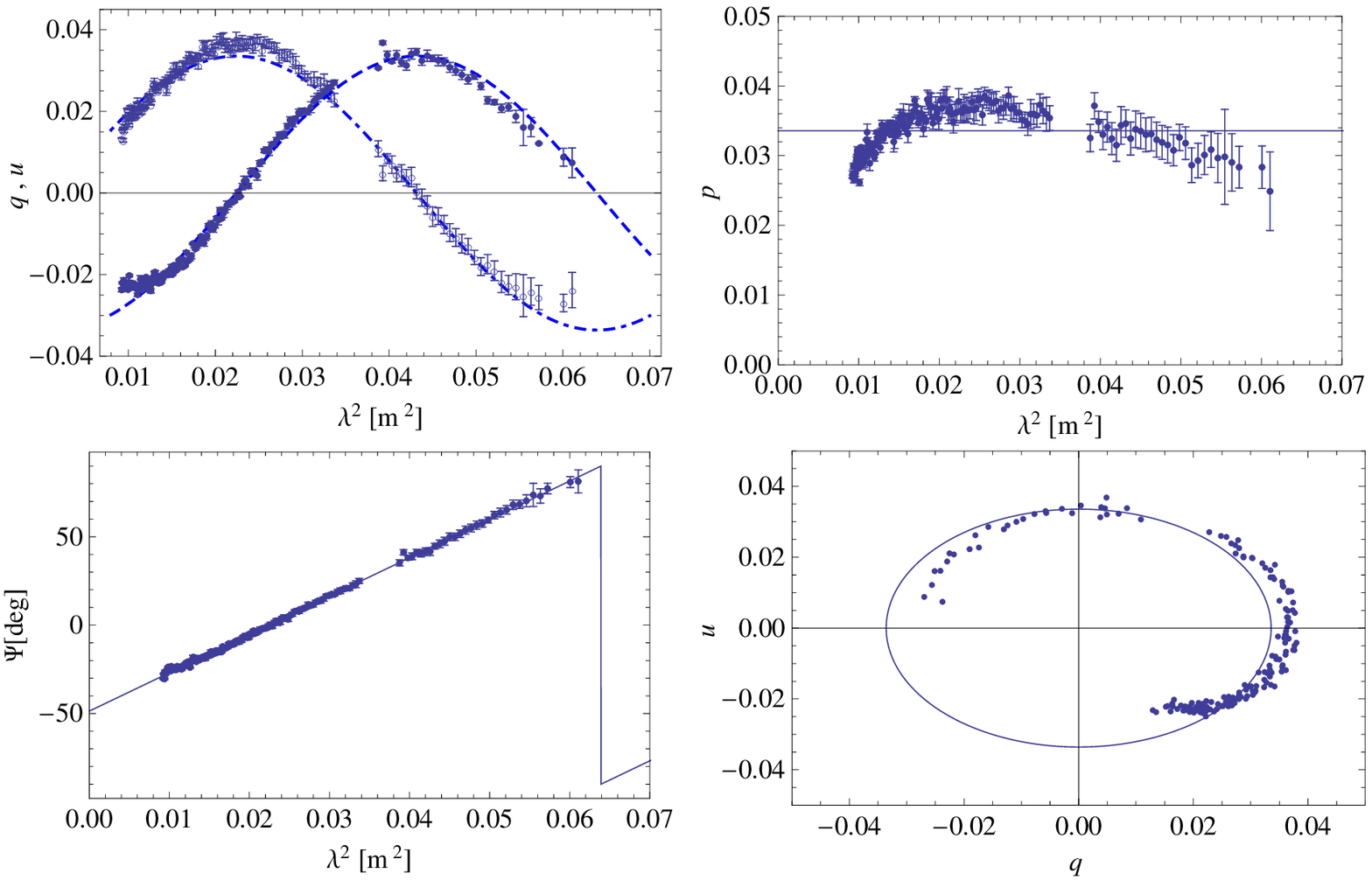}
  \caption{Polarization data for PKS B0454-810, and the 
  corresponding best-fit single RM-component model (Eqn.~\ref{simple}). 
  Layout as described in Figure~\ref{1903simple}. 
}
  \label{0454simple}
\end{figure*}

\begin{figure*}
     \includegraphics[width=14cm]{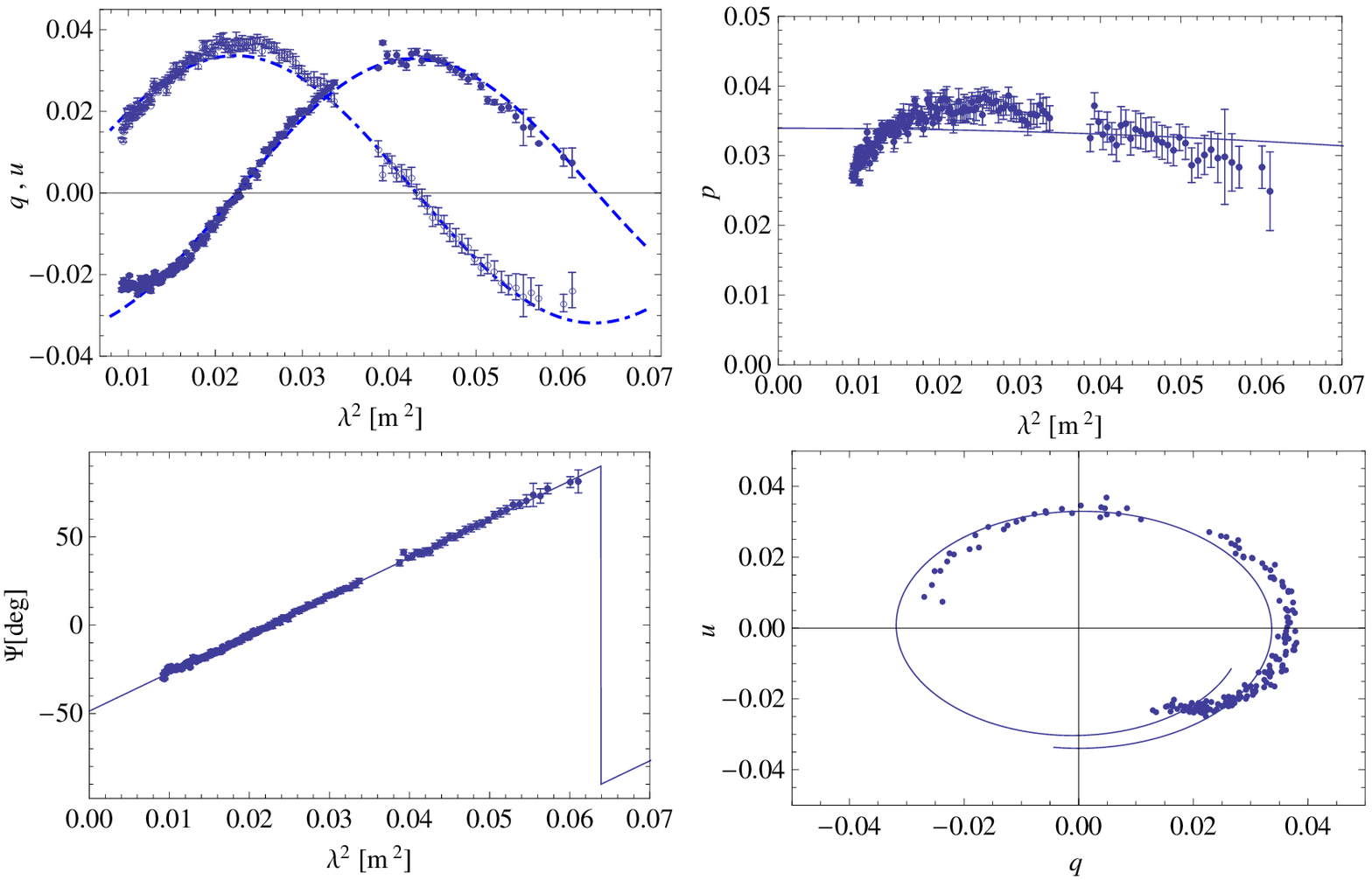}
  \caption{As for Figure~\ref{0454simple} for PKS B0454-810, but fit 
  by a single RM-component model with depolarization from 
  external Faraday dispersion (Eqn.~\ref{EFD}). 
}
  \label{0454depol}
\end{figure*}

\begin{figure*}
     \includegraphics[width=14cm]{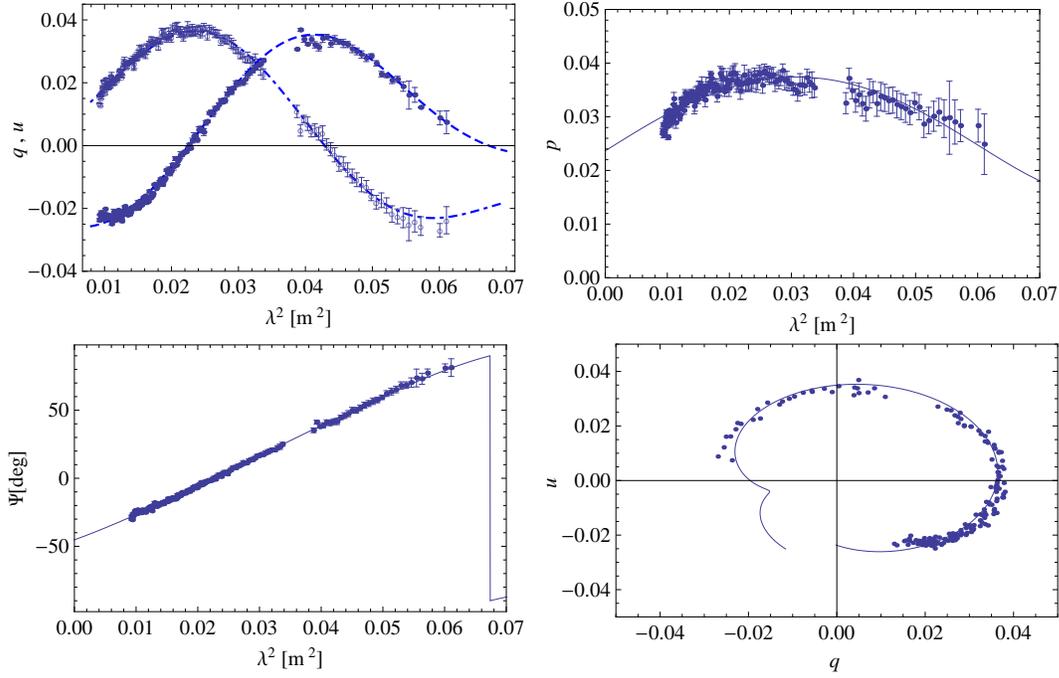}
  \caption{As for Figure~\ref{0454simple} for PKS B0454-810, 
  but fit with a two RM-component model. The kink in the $q(\lambda^2)$ 
  vs.~$u(\lambda^2)$ plot is a unique signature of this particular model. 
}
  \label{0454_2cmpnt}
\end{figure*}

As can be seen in Figures~\ref{0454simple} \& \ref{0454depol}, both single-component RM models 
(with and without a depolarizing screen) provide poor fits to the data. 
They both determine approximately the same RM ($+37.8\pm0.2$~rad~m$^{-2}$) 
but cannot explain the observed decrease 
in the degree of polarization towards the shortest wavelengths. 
The two-component model, shown in Figure~\ref{0454_2cmpnt}, does much better at 
describing the $p(\lambda^2)$ data while also providing a good description 
of $\Psi(\lambda^2)$. We note that the RM of the strongest polarized component is now 
significantly different ($+29.2\pm1.1$~rad~m$^{-2}$) from that found in the one-component 
models or from the peak in Faraday depth inferred from RM synthesis in 
Figure~\ref{RMCLEAN}(b). 

While both the reduced chi-squared and BIC values strongly favour the two RM-component 
model over the single component models (Table~3), 
on closer inspection it is clear that the $p(\lambda^2)$ data 
at the shortest wavelengths observed are not 
very well fit by the two RM-component model either (Fig.~\ref{0454_2cmpnt}). 
Hence, we conclude that the data are not well fit by any of the three models listed 
in Table~3. 

We consider a possible alternative explanation for this source in terms of 
polarization propagation effects as a function of optical depth within the source. 
Specifically, the case of one optically thick and one 
optically thin component; the optically thin and strongly polarized component would dominate 
the emission at the longer wavelengths ($\lambda^2>0.025$~m$^2$) while the observed 
depolarization could be explained by external Faraday dispersion. At the shorter 
wavelengths ($\lambda^2<0.025$~m$^2$), the optically thick and weakly polarized 
component would become more dominant coupled with the optically thin component 
becoming fainter leading to a decrease of the observed degree of polarization in 
a non-trivial manner. This idea is supported by the inverted spectrum indicative of 
a synchrotron self-absorbed region caused by either multiple, discrete spectral components 
or a smooth distribution of magnetic field and electron density along the jet. 
If this model were correct, then we would expect at even shorter wavelengths, 
as the emission spectrum turns over, the degree of polarization should increase again.
However, detailed modelling of the polarized radiative transfer from such a region is 
required for a quantitative analysis and we defer such a study for a later paper.

\subsection{PKS B1610-771}

\begin{figure*}
    \includegraphics[width=14cm]{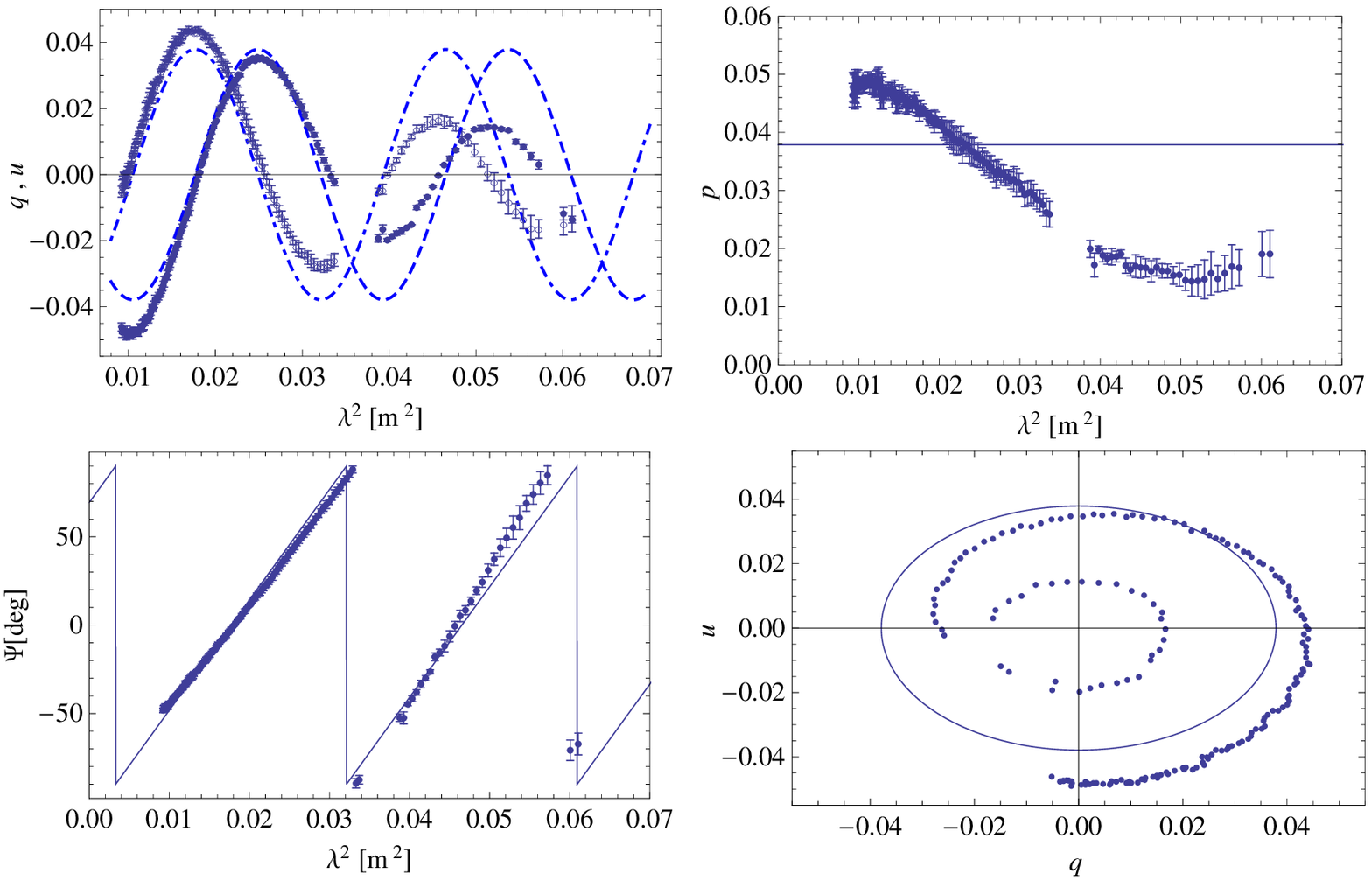}
  \caption{Polarisation data for PKS B1610-771, and the corresponding 
  best-fit single RM-component model (Eqn.~\ref{simple}). 
  Layout as described in Figure~\ref{1903simple}. 
}
  \label{1610simple}
\end{figure*}

\begin{figure*}
    \includegraphics[width=14cm]{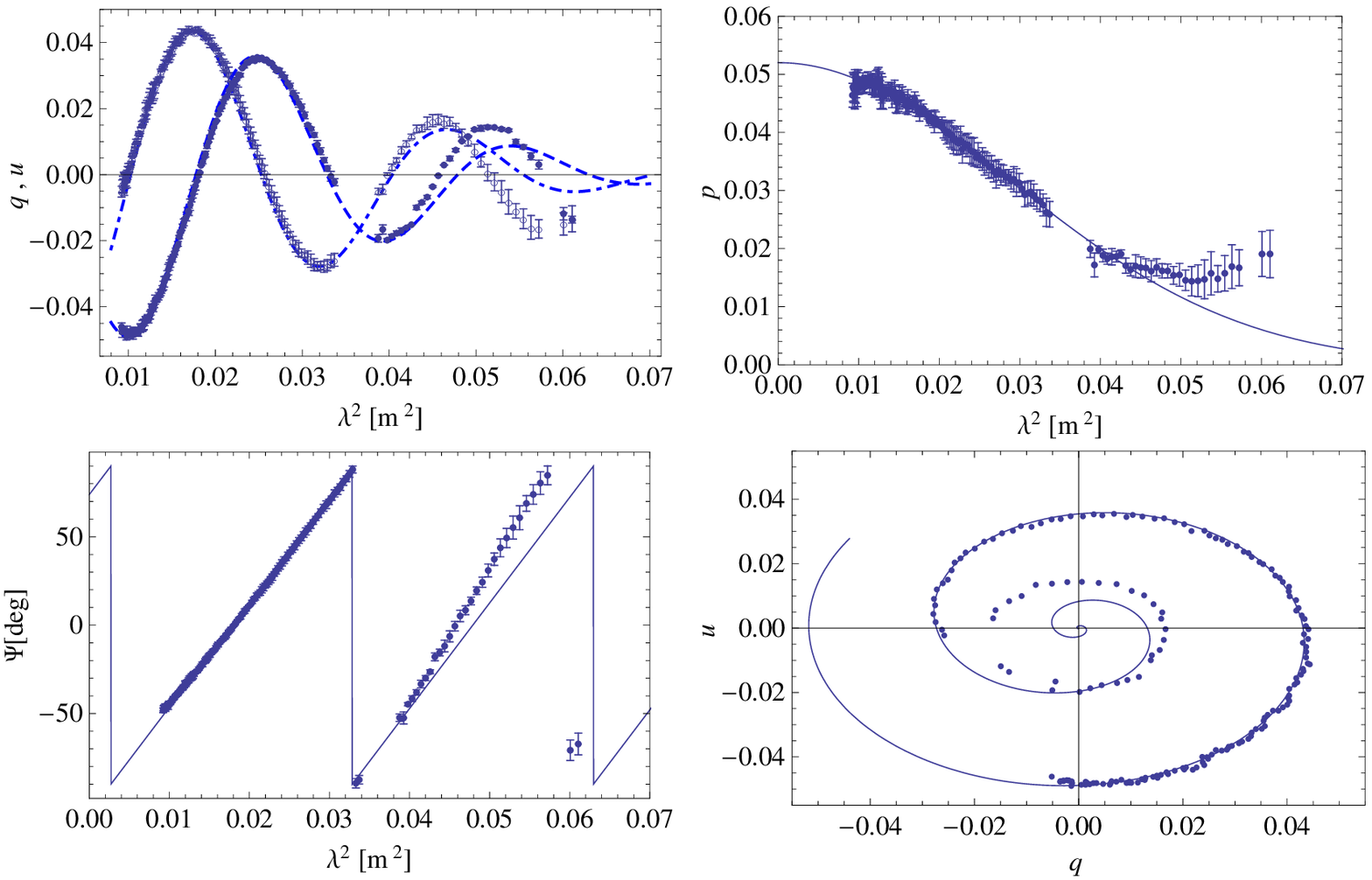}
  \caption{As for Figure~\ref{1610simple} for PKS B160-771, but fitted by a single 
  RM-component model with depolarization from external Faraday dispersion (Eqn.~\ref{EFD}). 
}
  \label{1610depol}
\end{figure*}

Figure~\ref{1610simple} shows how a simple one-component model provides a very poor fit to the 
complex polarization data for this source. 
A depolarizing screen model, shown in Figure~\ref{1610depol}, gives a good fit at the short wavelengths 
but fails to adequately fit the long wavelength data. 
These plots further highlight two interesting features of the data. First, it is clear that the slope of the 
$\Psi(\lambda^2)$ relationship gets steeper towards longer wavelengths and, second, 
while the $p(\lambda^2)$ plot shows that emission is strongly 
depolarized it also shows evidence for a reversed trend of increasing polarisation with 
$\lambda^2$ at longer wavelengths. 
Any simple depolarization model cannot explain both these effects. 

\begin{figure*}
    \includegraphics[width=14cm]{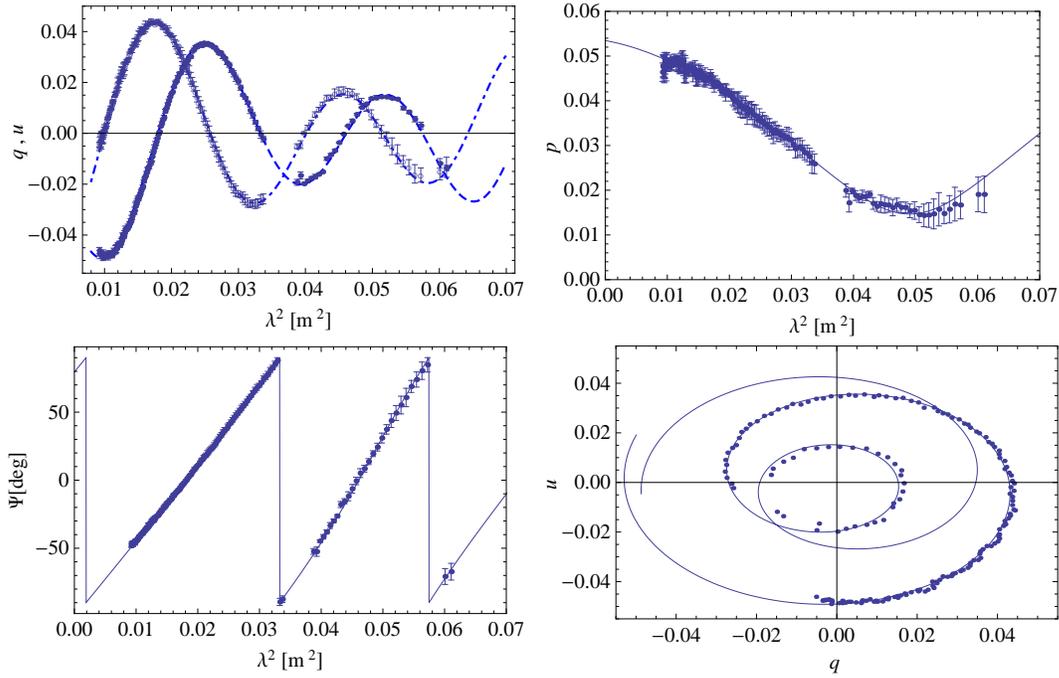}
  \caption{As for Figure~\ref{1610simple} for PKS B160-771, but fitted with a 
  two RM-component model. 
}
  \label{1610_2cmpnt}
\end{figure*}

A two RM-component model, shown in Figure~\ref{1610_2cmpnt}, provides a much better description 
of the data as demonstrated by the corresponding $\chi^2_{\nu}$ value of 1.04. 
The fit accounts for both the changing slope of $\Psi(\lambda^2)$ as well as 
the increasing $p(\lambda^2)$ for $\lambda^2 > 0.05$~m$^2$. 
The best-fit RMs are $+107.1\pm0.2$~rad~m$^{-2}$ and $+78.7\pm0.4$~rad~m$^{-2}$ 
for the first and second component, respectively. 
The BIC strongly favours the two RM-component model over the depolarizing screen 
model (see Table~3 for values).

\subsection{PKS B1039-47}
\begin{figure*}
    \includegraphics[width=14cm]{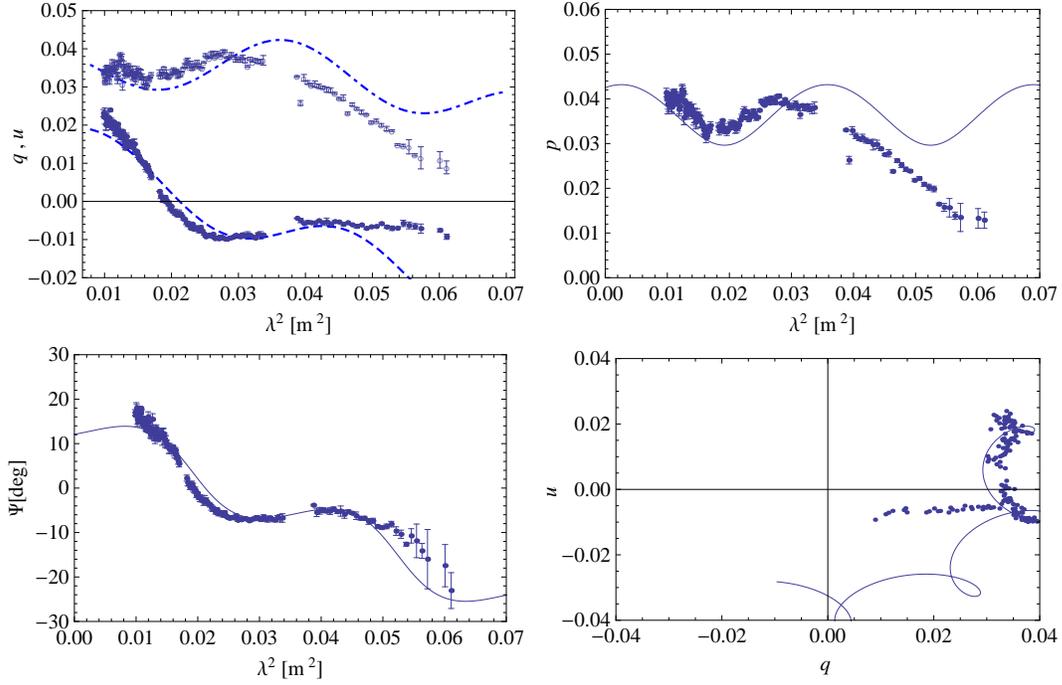}
  \caption{Polarisation data for PKS B1039-47, and the corresponding 
  best-fit two RM-component model. Layout as described in Figure~\ref{1903simple}. 
}
  \label{1039_2cmpnt}
\end{figure*}

This is the most striking source in terms of complex polarization structure. Figure~\ref{1039_2cmpnt} shows 
how both $\Psi(\lambda^2)$ and $p(\lambda^2)$ display non-linear, oscillatory behaviour indicative 
of multiple RM components. The RMCLEAN spectrum also indicates the presence of 
multiple RM components (Fig.~\ref{RMCLEAN}d). 

We list a single RM-component model in Table~3 for completeness but do not show 
the fit. A two RM-component model (both components Faraday thin) also provides a poor description of the 
data (Fig.~\ref{1039_2cmpnt}) with a reduced-$\chi^2$ value of 14, and this can be seen quite clearly 
in the $p(\lambda^2)$ distribution. 
We then tried models with one Faraday thin and one Faraday thick component where 
the Faraday thick component was described by either Eqn.~\ref{DFR} or Eqn.~\ref{IFD}. In 
both cases, neither model converged to an acceptable solution and they are not shown here. 

Our next approach was to fit models with three RM components, which in the case of all 
Faraday thin components have a total of nine parameters (i.e. $p_0$, RM, \& $\Psi_0$ 
for each component). 
In order to find a good model, we first fixed the parameters of 
the dominant RM component taken from RMCLEAN and let the other 
six model parameters vary. We then used these results as input guesses for 
the final nine parameter model fit. 
This returned best-fit RMs of $-13.1\pm1.5$~rad~m$^{-2}$, $-29.8\pm2.4$~rad~m$^{-2}$ and 
$+68.4\pm1.6$~rad~m$^{-2}$, listed in order of highest to lowest polarized fractions for the 
individual components. This provides a good fit, shown in Figure~14, with a $\chi^2_{\nu}$ value of 1.2 (Table~3). 

We also tried different combinations of Faraday thin and thick components 
within a three RM-component model with the best model shown in Figure~15. 
Both the BIC and $\chi^2_{\nu}$ values are almost identical for both types of 
three-component model listed in Table~3, so they do not help us to clearly 
discriminate between models in this case. More data, at longer wavelengths, 
is required to determine which three-component model provides a better fit. 
A four-component model (all Faraday thin) was also tried but did not improve the 
fit.

\section{Discussion}
\subsection{Physical origin of the RM components}

We have conclusively shown in the previous section that for two sources (PKS B1610-771 \& PKS B1039-47) 
we can spectrally resolve multiple polarized components of spatially unresolved AGN. 
We now discuss the likelihood of these additional RM components coming from regions 
of polarized emission 
along the line of sight or from multiple polarized regions on the plane of the sky but 
within our synthesised beam. 
Polarized diffuse Galactic emission \citep{testori2008} and polarized emission from 
radio halos/relics in galaxy clusters \citep[e.g.][]{ferrari2008} are the most likely candidates for 
any additional polarized emission components along the line of sight. However, both these possibilities are 
highly unlikely for our particular observations since first, we do not have the sufficient 
short $uv$-spacings to detect the smooth Galactic emission and secondly, none of the sources studied 
here have any diffuse X-ray emission associated with them in the ROSAT All Sky 
Survey \citep[RASS;][]{voges1999}, which effectively rules out the presence of any 
significant emission from galaxy clusters along the line of sight. 

If the sources were spatially unresolved, double-lobed radio galaxies such that our 
line of sight to one of the lobes travelled through a different magneto-ionic medium, 
then the polarized emission from each lobe would experience different amounts 
of Faraday rotation and could show up in our data as two distinct RM components 
\citep[e.g.][]{slysh1965, goldsteinreed1984}. For blazar type sources we require 
variation in the magneto-ionic medium along the jet because we only 
detect the Doppler boosted emission from the jet orientated toward us. Many high-resolution studies 
of such objects have shown substantial variations in both RM and polarized intensity 
on parsec-scales \citep[e.g.][]{zt2003, zt2004, osullivangabuzda2009, hovatta2011}. 
Below we discuss what is already known for each source and why the mostly likely 
origin of the additional RM components is from the compact inner regions of the radio source. 

PKS B1903-802 is a flat spectrum radio quasar which, from 1.4 to 20 GHz, maintains a total flux 
density of $\sim$1~Jy while remaining polarized at $\sim$3\% \citep{MurphyAT20G}. 
From 43~GHz observations with the ATCA, we know that this source is unresolved down 
to at least 0.15" (Rajan Chhetri, private communication) which 
corresponds to a linear scale of $\sim$0.9~kpc for the quoted redshift of this source (Table~\ref{obs_params}). 
The source is resolved into two total intensity components on scales less than 5~mas (30~pc) 
from observations at 8.4~GHz with the Australian Long Baseline Array \citep[LBA;][]{ojha2005}. 
No spectral index or polarization information is available on these scales but at least one of 
these components must be the origin of the strong polarized emission we see in our data. 
Hence, it is likely that there is a significant contribution to the observed RM from the 
immediate environment of the source and its host galaxy as well as the Faraday rotation 
caused by our own Galaxy. 

PKS B0454-810 is a well studied flat spectrum radio quasar \citep[e.g.][]{ricci2004} 
and has an inverted radio light curve which begins to turn over above 100~GHz 
\citep{wmap2003}. \cite{ricci2004} detected polarized emission of $\sim$2.6\% at 18.5~GHz 
while \cite{MurphyAT20G} quote an upper limit of $1.5\%$ from 20~GHz ATCA observations. 
While it has been detected in X-rays \citep{voges1999}, it does not have a $\gamma$-ray detection 
from Fermi \citep{fermi1, fermi2}. It is spatially unresolved down to at least 5" ($\sim$30~kpc) from 
inspection of the visibilities on the longest baselines in our observations. It has also been imaged 
on milliarcsecond scales with both the VLBI Space Observatory Programme \citep{dodson2008} 
and the LBA \citep{ojha2005}, showing that it is resolved on scales less than 5 mas. 
\cite{Rayner2000} found the source to be circularly polarized, with a $>10\sigma$ detection 
at both 1.4 and 5~GHz, which is indicative of a core-dominated AGN. Therefore, all this data 
supports our assertion that the linearly polarized emission we detect is coming 
from the compact inner regions of the AGN jet and provides some weight to our 
explanation for the observed variation in $p(\lambda^2)$ being due to the combination 
of a weakly polarized optically-thick region and a strongly polarized optically-thin region.

PKS B1610-771 has been extensively studied across a wide range of wavelengths. 
It is classified as a flat-spectrum radio quasar \citep[e.g.][]{healey2007}, is significantly 
polarized (1--2\% level) at 5, 8 and 20~GHz \citep{massardi2008} and is 
highly polarized ($>$3\%) in the optical \citep{veroncetty2006}. 
It is coincident with an unresolved X-ray source \citep{voges1999} and has 
a GeV $\gamma$-ray detection in both the first and second Fermi-LAT catalogues \citep{fermi1, fermi2}. 
No source structure is detected on scales greater than 0.15" (Chhetri, private comm.), which corresponds 
to a linear scale of $\sim$1.3~kpc at its redshift of 1.71 \citep{hunstead1980}. 
PKS B1610-771 is also seen to exhibit interstellar scintillation at low radio frequencies, 
with a characteristic time scale of 400 days \citep{gaenslerhunstead2000}. This further 
supports the conclusion that its flux is dominated by compact rather than extended components.
On milliarcsecond scales at 8.4~GHz the jet extends in a North-West direction with several 
bright jet knots in total intensity seen out to a projected distance of $\sim$130 pc \citep{ojha2010}. 
From our analysis of this source, we predict that two of these knots are strongly polarized 
and have different RMs. 

PKS B1039-47 is classified as a flat-spectrum radio quasar that is located 
along a sightline $\sim$10$^{\circ}$ from the Galactic plane, and has been 
measured to be $\sim$4\% polarized at 20~GHz \citep{massardi2008}. 
The host galaxy has a measured redshift of 2.59 (O. Titov, private communication)
and there is no X-ray or $\gamma$-ray detection for this source. 
The emission structure remains unresolved down to 0.15" (1.2~kpc), and an LBA 
image from \cite{ojha2004} shows a jet extending to the North-West 
out to $\sim$20~mas (160 pc), composed of three bright total intensity regions. Our 
analysis in this paper, which finds a best-fit three RM-component model for 
this source, suggests that each of these regions is polarized. 

Therefore, in the case of PKS B1610-771 and PKS B1039-47, we claim to have 
detected separate polarized components in the compact inner regions of the jet 
on parsec-scales that are illuminating an inhomogeneous magneto-ionic medium 
in the immediate vicinity of the jet. The largest RM difference between different 
components that we have found from our model fits is $\sim$100~rad~m$^{-2}$ 
(between the second and third components in PKS B1039-47). 
This is not inconsistent with recent measurements on milliarcsecond scales at 
1.4~GHz where variations of 10s to 100s of rad~m$^{-2}$ have been 
measured in several blazars \citep{coughlan2011}. 
We can test our predictions for these sources through multi-frequency polarization 
sensitive observations with the LBA, which will allow us to 
spatially probe their milliarcsecond scale polarized structure for the first time.

\subsection{RM time-variability}
If we attribute the RM difference between components to the magneto-ionic material 
in the immediate vicinity of the parsec-scale jet then there are obvious implications for any 
observed time-variability from RM measurements on arcsecond scales and greater. 
The polarized and RM structure of VLBI jets has been observed to vary on timescales 
as short as months \citep[e.g.][]{gomez2011, zt2001}, so the relative polarized flux of 
the individual components detected in our ATCA observations may also vary on similar timescales. 
Therefore, the relative 
difference between the individual component RM values may also vary if the polarized components 
are moving along the jet and illuminating different parts of an inhomogeneous Faraday 
screen close to the AGN.
This means that RM time-variability for observations on similar angular scales to those 
presented in this paper may be simply due to the observations sampling different 
dominant RM components as they move along the jet 
or, if the observations do not have sufficient frequency coverage and spectral resolution, 
a complicated combination of multiple RM components that does not 
accurately represent any of the components. 

\cite{law2011} were able to compare four sources from their low spatial resolution, 
wide-bandwidth 1--2~GHz observations with high spatial resolution RM maps from 
the Very Long Baseline Array (VLBA). In general, they did not detect the high fractional 
polarization or high RM values seen from 5 to 22~GHz in the VLBA images. 
This is not surprising since the high spatial resolution of VLBA observations are less 
affected by beam depolarization and also because RMs on these 
scales have been observed to increase with increasing frequency \citep{osullivangabuzda2009}. 
Another possibility is that there may be intermediate scale polarized structure 
that the VLBA is not sensitive to, negating the validity of the comparison. 
Therefore, parsec-scale VLBI observations at similar frequency ranges 
taken as close in time as possible to the low spatial-resolution, 
wide-bandwidth observations are required for direct comparison. 
If the results from both these types of observations can be linked then it may be possible 
with multi-epoch polarization observations with wide-bandwidth facilities like the ATCA 
to map out the parsec-scale evolution of polarized components 
as well as the Faraday rotating environment in AGN jets, as suggested by \cite{law2011}. 

Even though the majority of the Faraday rotation occurs as the polarized radiation passes 
through our Galaxy, the RM difference between multiple components as well as any observed 
variability is likely due to the magneto-ionic material in the immediate vicinity of the AGN 
jet. 
Hence, the type of sources studied in this paper may not be suitable as primary 
polarization calibrators since the values for their RMs and degree of polarization are 
likely to change on short timescales. A more stable type of calibrator source would be 
one in which the polarized emission comes from extended emission regions such as 
the lobes which vary on much longer timescales. 

\subsection{Reliability of RMCLEAN}
For PKS B1903-802 the peak RM found using RMCLEAN agrees very well 
(within $\sim$0.3~rad~m$^{-2}$) with the mean RM found from our best-fit 
external Faraday dispersion model. 
In the case of PKS B0454-810, the RM extracted using RMCLEAN differs 
by $\sim$2~rad~m$^{-2}$ from what we consider to be the correct RM 
for this source. However, the comparison in 
this case is somewhat unreliable since we have not been able to conclusively 
identify the correct polarization model for this source. 

There have been some questions raised in the literature about the ability of the RMCLEAN 
method to accurately recover multiple RM components for sources with complex Faraday 
structure \citep[e.g.][]{frick2010, farnsworth2011}.  
We find that the RMCLEAN method performs poorly in recovering the correct RMs for 
PKS B1610-771 and PKS B1039-47. While it does predict the presence of multiple 
RM components, the clean-component distribution does not match what is 
found through model-fitting $q(\lambda^2)$, $u(\lambda^2)$ (Fig.~\ref{RMCLEAN}c,d). 
For PKS B1039-47, we investigated whether or not the three best-fit model RM components 
could be recovered with the same $\lambda^2$-coverage but with no noise. 
Figure~\ref{1039model} shows that again, after RMCLEAN, the clean-component distribution 
does not associate polarized power with the correct RM model-component locations. 
On reflection this may not be very surprising given that the RM resolution of our 
experiment is $\sim$60~rad~m$^{-2}$ and is therefore unable to resolve multiple 
RM components which differ by less that this value. 
Thus, in cases of complex RM structure of extragalactic point sources, 
alternative reconstruction algorithms \citep[e.g.][]{li2011} need to be 
investigated for application in all-sky RM surveys. 
Currently, the best approach for sources with complex Faraday depth structure 
is model-fitting multiple RM-component models to the observed $q(\lambda^2)$ and 
$u(\lambda^2)$ in order to find the most likely physical model for the source.

\begin{table}
 \caption{Comparison between narrow-band and wide-band RMs}
 \centering
   \begin{tabular}{cccccc}
    \hline\hline
 Source &      RM$_{\rm wide}$     & RM$_{\rm old~ATCA}$  & RM$_{\rm NVSS}$ \\
               &         [rad~m$^{-2}$]           & [rad~m$^{-2}$]    & [rad~m$^{-2}$] \\
    \hline
      PKS B1903-802  &       $+18.1\pm0.1$    &     $+18.2\pm0.4$ & $+21.5\pm1.3$ \\
      PKS B0454-810  &       $+37.8\pm0.2$    &     $+39.9\pm1.2$ & $+38.9\pm1.5$ \\
      PKS B1610-771  &       $+107.1\pm0.2$  &     $+134.6\pm1.0$ & $+128.5\pm2.7$ \\
      PKS B1039-47    &       $-13.1\pm1.5$      &     $-13.0\pm6.7$ & $-8.9\pm1.2$ \\
      \hline
   \end{tabular}
   \scriptsize{RM$_{\rm wide}$: RM of main component derived from 1.1--3.1~GHz data. 
   RM$_{\rm narrow}$: RM derived using data from 1304--1494~MHz.
   RM$_{\rm NVSS}$: derived using the same frequency coverage as in \cite{taylor2009}. }
\label{RMcomparison}
\end{table}

\subsection{Some implications for RM surveys}
Taking the same data but restricting the frequency range to 20 x 10~MHz channels, 
centred at 1.4~GHz, allowed us to compare the RMs derived from the full 2~GHz 
bandwidth with the previous narrow-band system on the ATCA \citep[e.g.][]{feain2009}. 
Using the fitting procedure employed throughout this 
paper, we find that we get the same RMs (within the errors) using 200~MHz of data for both PKS B1903-802 
and PKS B0454-810. In this case, a single RM-component model with external Faraday dispersion 
still provides a better fit over a model without depolarization. 
In Table~\ref{RMcomparison}, for each source, we list the RM of the strongest polarized 
component derived from the full 1.1--3.1~GHz data to compare with the RM derived 
from the restricted range of 1.340--1.494~GHz.

For PKS B1610-771 we obtain a good fit ($\chi^2_{\nu}=1.2$, BIC$=-475$) for a single RM-component 
with external Faraday dispersion giving a mean RM of $+134.6\pm1.0$~rad~m$^{-2}$ and a dispersion 
of $10.9\pm0.5$~rad~m$^{-2}$. 
A marginally poorer fit is found using a two RM-component model ($\chi^2_{\nu}=1.3$, BIC$=-452$) 
with RM$_1=+110.1\pm1.8$~rad~m$^{-2}$ and RM$_2=+79.7\pm2.1$~rad~m$^{-2}$. 
Therefore, in this case we would adopt the single RM-component model since the evidence does 
not favour the more complex two RM-component model. But we already know that the two 
RM-component model is preferred using the full 2~GHz bandwidth data. 
The difference in RM between the single-component model and 
the strongest component in the two RM-component model is $\sim25$~rad~m$^{-2}$ (Table~\ref{RMcomparison}). 
This is quite a dramatic example of how wrong one can be in estimating the RM using 
narrow-bandwidth data.  
This highlights the important role wide-bandwidth data 
play in determining the correct Faraday depth structure of AGN on these angular scales.  
The best-ft model for PKS B1039-47 using the data from 1.340--1.494~GHz has 
two RM components where the strongest component has an RM equal (within the errors) 
to that found from the 1.1--3.1~GHz data. 

Due to our small sample size we are unable to comment on 
how often additional RM components may be detected in extragalactic sources. 
From a sample of 37 bright, polarized sources \cite{law2011} found that 
$\sim$25\% of sources had an extra RM component detected with a significance greater 
than 7$\sigma$ ($\sim$40~mJy). 
They also found that the polarized flux weighted mean RM was similar to the low resolution 
RMs quoted by \cite{taylor2009}. This suggests that sources like PKS B1610-771 may 
be rare where the flux weighted mean RM ($\sim$97~rad~m$^{-2}$) is significantly 
different from the RM which is derived from narrow-bandwidth observations. 
We also list in Table~\ref{RMcomparison} the RMs calculated from our data using the same 
frequency setup as used by \cite{taylor2009}. For the simple sources, the RMs 
agree within 2--3~rad~m$^{-2}$. 

Current and upcoming spectropolarimetric all-sky RM surveys such as GALFACTS \citep{galfacts} and 
the POSSUM survey on ASKAP \citep{possum} plan to extract RMs from observations 
with 300~MHz of instantaneous bandwidth near 1.4~GHz. 
For sources with a single RM-component modified by depolarization from external Faraday dispersion, 
observations with 300~MHz of bandwidth will produce the same results as we have found 
using 2~GHz of bandwidth (e.g. PKS B1903-802). 
In the case of sources with multiple RM components, it is strongly recommended that modelling of 
$q(\lambda^2)$, $u(\lambda^2)$ be undertaken instead of using the reconstruction algorithm RMCLEAN. 
Since ASKAP will have much better RFI conditions than at the ATCA, 
we use simulated data from the best-fit models for PKS B1610-771 
and PKS B1039-47 (highlighted in bold in Table~3) instead of the observed data. 
Using the planned POSSUM frequency coverage of 1130--1430~GHz, we find that our modelling 
procedure would recover the correct two RM-component model for PKS B1610-771. 
For PKS B1039-47, 
depending on the quality of the data, it may be difficult to determine 
whether a two or three RM-component model provides the best fit.  
Complementary observations from a 300~MHz band at lower frequencies (e.g. 0.7--1.0~GHz) 
would enable us to recover the correct simulated three RM-component model in this case.

\section{Conclusion}

Using the new wide-bandwidth receivers on the ATCA, we have shown
that we can spectrally resolve the polarization structure of spatially 
unresolved radio sources. 
We have identified two AGN (PKS B1610-771 and PKS B1039-47) 
where more than one rotation measure (RM) component is required to 
describe the Faraday structure of the source. 
We further demonstrate that modelling of both the polarization angle and 
degree of polarization dependences with wavelength squared is 
essential in determining the true Faraday depth structure of extragalactic 
point sources. 
We also find that the RM synthesis reconstruction algorithm RMCLEAN does not 
recover the correct RMs for sources with multiple RM components in our data. 

The most likely origin for the additional RM components in both PKS B1610-771 
and PKS B1039-47 is from the compact inner jet regions on parsec scales. 
This leads us to suggest that RM time-variability in extragalactic point sources 
may be due to the evolving polarized jet structure on parsec scales which 
illuminates different parts of an inhomogeneous magneto-ionic medium in the 
immediate vicinity of the jet. 
Follow-up observations of these particular sources with parsec-scale spatial 
resolution using the Australian Long Baseline Array (LBA) will enable us to 
test our predictions.  
Hence, with multi-epoch polarization observations using wide-bandwidth facilities 
like the ATCA it may be possible to map out the parsec-scale evolution of polarized 
components as well as the Faraday rotating environment in AGN jets. 

In the near future, combining data from the pristine RFI environment of ASKAP from 
0.7 to 1.8~GHz with data from the ATCA at higher frequencies can provide an exquisite probe 
of the polarization properties of a much larger sample of AGN.

\clearpage

\begin{figure*}
    \includegraphics[width=14cm]{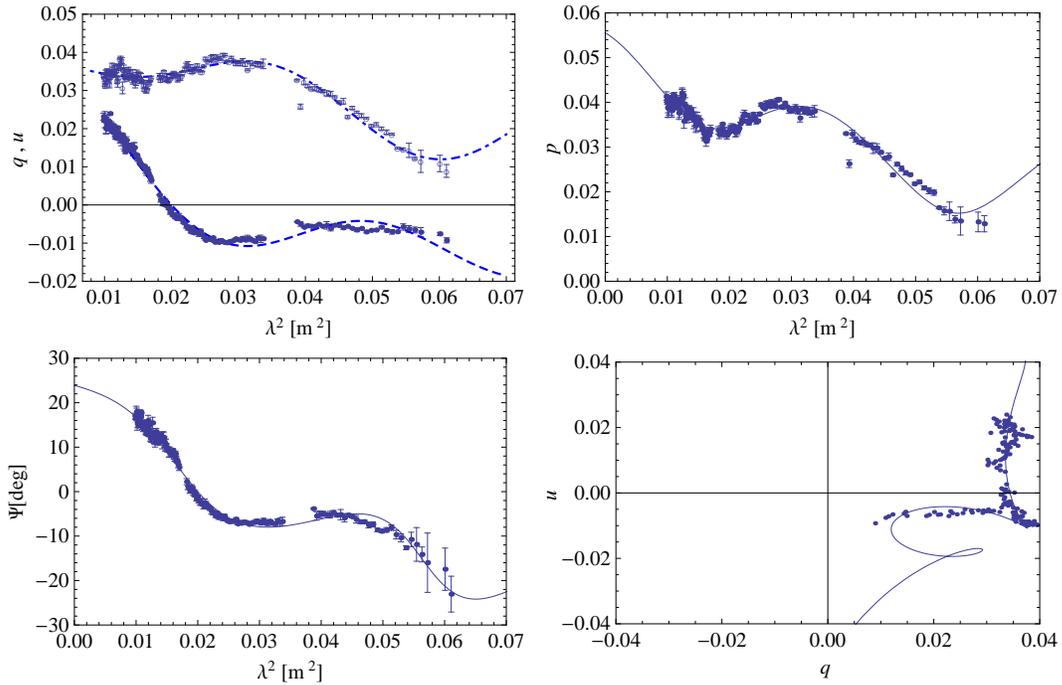}
  \caption{As for Figure~\ref{1039_2cmpnt} for PKS B1039-47, 
  but fit by a three RM-component model (all Faraday thin). 
}
  \label{1039_3cmpnt}
\end{figure*}


\begin{figure*}
    \includegraphics[width=14cm]{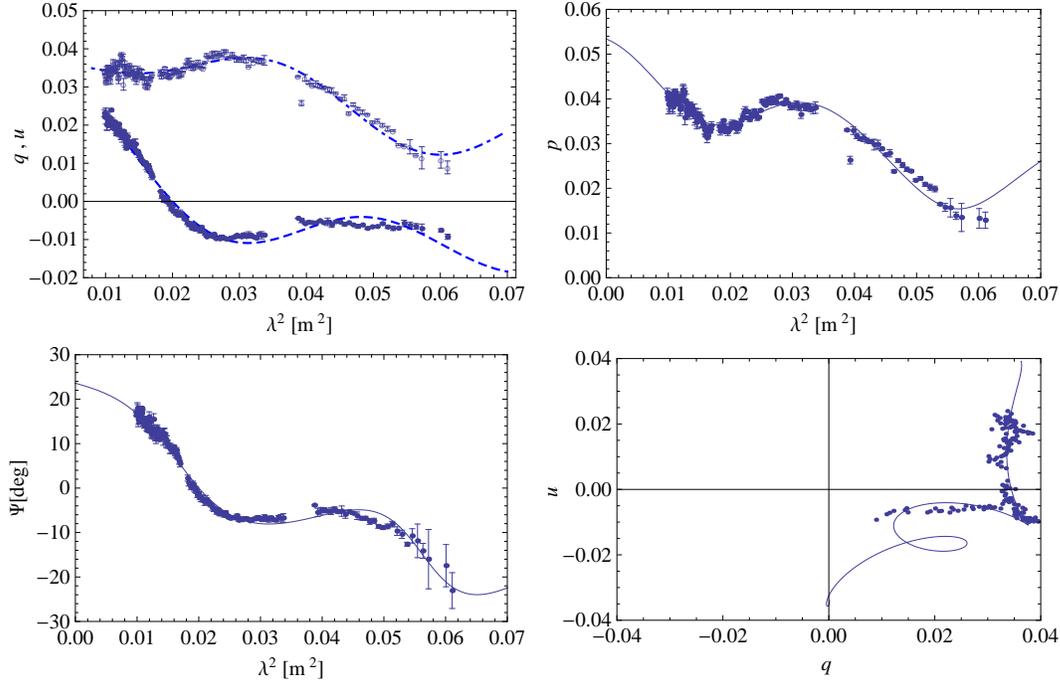}
  \caption{As for Figure~\ref{1039_2cmpnt} for PKS B1039-47, 
  but fit by a three RM-component model (two Faraday thin, 
  one Faraday thick: Eqn.~\ref{DFR}). 
}
  \label{1039_2thin1thick}
\end{figure*}

%
\begin{figure}
    \includegraphics[width=8cm]{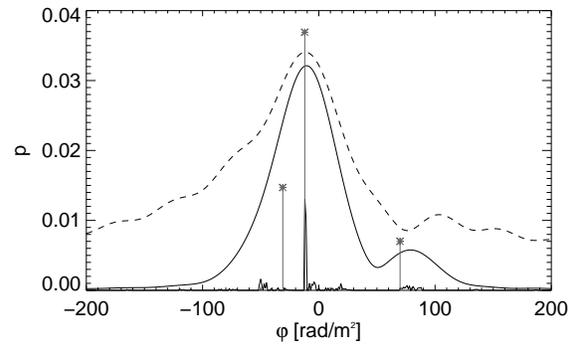}
  \caption{Simulated three component model without noise for PKS B1039-47 but 
  with identical $\lambda^2$ coverage as in Figure~\ref{RMCLEAN}(d).
  Stars: input model components taken from the best fit to PKS B1039-47 data 
  shown in Figure~\ref{1039_3cmpnt}.
  Dashed line: Dirty RM synthesis spectrum from simulated data. 
  Solid line: RMclean spectrum from simulated data. 
  Clean-component positions and amplitudes are also included on plot. 
}
  \label{1039model}
\end{figure}

\section{Acknowledgements}
The Australia Telescope Compact Array is part of the Australian Telescope, which is funded by the Commonwealth of Australia for operation as a National Facility managed by CSIRO.
We would like to thank the engineers, technicians and staff at CSIRO's Marsfield and Narrabri sites who were involved in the successful upgrade of the 20/13 cm receiver systems to 
take advantage of the 2~GHz correlator bandwidth available in CABB. 
B.M.G. and T.R. acknowledge the support of the Australian Research Council through grants DP0986386 and FS100100033, respectively. 
S.P.O'S. would like to thank David McConnell, Jamie Stevens, Mark Weiringa, Julie Banfield, Tim Cawthorne, Russell Jurek and Chris Hales for helpful discussions. 
This research has made use of NASA's Astrophysics Data System Service and the NASA/IPAC Extragalactic Database (NED) which is operated by the Jet Propulsion Laboratory, California Institute of Technology, under contract with the National Aeronautics and Space Administration. 

\bibliographystyle{mn2e}
\bibliography{cabb_submit}
\bsp

\end{document}